\documentclass[fleqn,usenatbib]{mnras}

\usepackage{newtxtext,newtxmath}

\usepackage[T1]{fontenc}

\DeclareRobustCommand{\VAN}[3]{#2}
\let\VANthebibliography\thebibliography
\def\thebibliography{\DeclareRobustCommand{\VAN}[3]{##3}\VANthebibliography}

\usepackage{graphicx}	
\usepackage{amsmath}	
\usepackage{amssymb}	
\usepackage{mydefs}     
\usepackage[dvipsnames]{xcolor}      
\usepackage{float} 
\usepackage{ulem}
\usepackage{xspace}

\newcommand\ABACUS{\textsc{abacus}\xspace}

\title[DESI-\textsc{FastPM}]{The DESI $N$-body Simulation Project -- II. Suppressing sample variance with fast simulations}

\author[Ding et al.]
{Zhejie Ding,$^{1,2}$\thanks{E-mail: zhejied@sjtu.edu.cn}
Chia-Hsun Chuang,$^{3,4}$\thanks{E-mail: albert.chuang@utah.edu}
Yu Yu,$^{1, 2}$\thanks{E-mail: yuyu22@sjtu.edu.cn}
Lehman H.~Garrison,$^{5}$ 
Adrian E.~Bayer,$^{6,7}$
Yu Feng,$^{7}$
\newauthor
Chirag Modi,$^{8,5}$ 
Daniel J.~Eisenstein,$^{9}$
Martin White,$^{6,7,10,11}$
Andrei Variu,$^{12}$
Cheng Zhao,$^{12}$
Hanyu Zhang,$^{13}$
\newauthor
Jennifer Meneses Rizo,$^{14}$
David Brooks,$^{15}$
Kyle Dawson,$^{4}$
Peter Doel,$^{15}$
Enrique Gaztanaga,$^{16,17}$
Robert Kehoe,$^{18}$
\newauthor
Alex Krolewski,$^{19,20}$
Martin Landriau,$^{11}$
Nathalie Palanque-Delabrouille,$^{11,21}$
Claire Poppett$^{22}$
\\
$^{1}$Department of Astronomy, School of Physics and Astronomy, Shanghai Jiao Tong University, Shanghai, 200240, China\\
$^{2}$Key Laboratory for Particle Astrophysics and Cosmology(MOE)/Shanghai Key Laboratory for Particle Physics and Cosmology, Shanghai, 200240, China\\
$^{3}$Kavli Institute for Particle Astrophysics and Cosmology, Stanford University, 452 Lomita Mall, Stanford, CA 94305, USA\\
$^{4}$Department of Physics and Astronomy, University of Utah, Salt Lake City, UT 84112, USA\\
$^{5}$Center for Computational Astrophysics, Flatiron Institute Simons Foundation, 162 Fifth Ave. New York, NY 10010, USA\\
$^{6}$Berkeley Center for Cosmological Physics, University of California, Berkeley, CA 94720, USA\\
$^{7}$Department of Physics, University of California, Berkeley, CA 94720, USA\\
$^{8}$Center for Computational Mathematics, Flatiron Institute, New York, NY 10010, USA\\
$^{9}$Center for Astrophysics | Harvard \& Smithsonian, 60 Garden St., Cambridge, MA 02138, USA\\
$^{10}$Department of Astronomy, University of California, Berkeley, CA 94720, USA\\
$^{11}$Lawrence Berkeley National Laboratory, 1 Cyclotron Road, Berkeley, CA 94720, USA\\
$^{12}$Institute of Physics, Laboratory of Astrophysics, Ecole Polytechnique F\'ed\'erale de Lausanne (EPFL), Observatoire de Sauverny, CH-1290 Versoix, Switzerland\\
$^{13}$Department of Physics, Kansas State University, Manhattan, KS 66506, USA\\
$^{14}$Instituto de F\'isica,  Universidad Nacional Aut\'onoma de M\'exico, Apdo. Postal 20-364, Ciudad de M\'exico, M\'exico \\
$^{15}$Department of Physics \& Astronomy, University College London, Gower Street, London, WC1E 6BT, UK\\
$^{16}$Institute of Space Sciences (ICE, CSIC), 08193 Barcelona, Spain\\
$^{17}$Institut d\'~Estudis Espacials de Catalunya (IEEC), 08034 Barcelona, Spain\\
$^{18}$Department of Physics, Southern Methodist University, Dallas, TX 75275, USA\\
$^{19}${AMTD Fellow, Waterloo Centre for Astrophysics, University of Waterloo, Waterloo ON N2L 3G1, Canada}\\
$^{20}${Perimeter Institute for Theoretical Physics, 31 Caroline St. North, Waterloo, ON NL2 2Y5, Canada}\\
$^{21}$IRFU, CEA, Universit\'e Paris-Saclay, Gif-sur-Yvette, F-91191, France\\
$^{22}$Space Sciences Laboratory (SSL), UC Berkeley, 7 Gauss Way, Berkeley, CA 94720, USA
}

\date{Accepted XXX. Received YYY; in original form ZZZ}

\pubyear{2022}

\begin{document}
\label{firstpage}
\pagerange{\pageref{firstpage}--\pageref{lastpage}}
\maketitle

\begin{abstract}
Dark Energy Spectroscopic Instrument (DESI) will construct a large and precise three-dimensional map of our Universe. The survey effective volume reaches $\sim20\Gpchcube$. It is a great challenge to prepare high-resolution simulations with a much larger volume for validating the DESI analysis pipelines.
\textsc{AbacusSummit} is a suite of high-resolution dark-matter-only simulations designed for this purpose, with $200\Gpchcube$ (10 times DESI volume) for the base cosmology.
However, further efforts need to be done to provide a more precise analysis of the data and to cover also other cosmologies.
Recently, the CARPool method was proposed to use paired accurate and approximate simulations to achieve high statistical precision with a limited number of high-resolution simulations.
Relying on this technique, we propose to use fast quasi-$N$-body solvers combined with accurate simulations to produce accurate summary statistics.
This enables us to obtain 100 times smaller variance than the expected DESI statistical variance at the scales we are interested in, e.g. $k < 0.3\hMpc$ for the halo power spectrum. In addition, it can significantly suppress the sample variance of the halo bispectrum.
We further generalize the method for other cosmologies with only one realization in \textsc{AbacusSummit} suite to extend the effective volume $\sim 20$ times.
In summary, our proposed strategy of combining high-fidelity simulations with fast approximate gravity solvers and a series of variance suppression techniques sets the path for a robust cosmological analysis of galaxy survey data.
\end{abstract}


\section{Introduction}
Dark Energy Spectroscopic Instrument (DESI) will collect more than 30 million galaxy spectra within 5 yr \citep{DESIscience2016}, constructing four main target catalogues, including bright galaxy sample, luminous red galaxies (LRGs), emission line galaxies (ELGs), and quasi-stellar objects. We will use them to determine the distances of the galaxies and construct a three-dimensional map of the large-scale structure of our Universe. 
A wealth of valuable information about our Universe is hidden in the statistics of the matter distribution, for example: 
the nature of dark energy \citep{Copeland_2006}, modified theories of gravity \citep{Huterer_2015,Alam_2020}, and neutrino mass \citep{LESGOURGUES_2006, DESInu, Allison_2015, hahn2020constraining, Kreisch_2021, Massara_2020, bayer2021fake,bayer2021detecting}.
To extract the cosmological information from observation, we need to build theoretical models that allow comparison to data. While the non-linear effect is difficult to model analytically, one alternative approach is running $N$-body simulations and thus building the models numerically. To this end, we want to generate simulations with huge volumes and high resolutions that are unfortunately limited by computing resources.

\cite{Grove2021} (the first paper of the DESI $N$-body simulation projects) compared multiple $N$-body codes and found good agreement which showed their robustness. Also, they compared the simulations with different mass resolutions and identified that the errors introduced by simulations with particle mass of $2.11\times10^{9}~h^{-1}$M$_{\sun}$ are well below the DESI statistical uncertainties. However, it is a challenge to generate simulations with sufficient volume for DESI with such a mass resolution. The physical survey volume of DESI is about $60\Gpchcube$ but the effective volume is about $20\Gpchcube$ if we take into account the observed galaxy densities at different redshifts \citep{DESIscience2016}. Thus, for the given mass resolution and volume, the simulation will need more than 800 billion particles. If we want to have the theoretical uncertainty below $1/10$ of DESI statistical error, the simulation volume should be 100 times larger, thus requiring 80 trillion particles. While it is not an impossible mission, we do not consider that it is a practical approach since it will cost a tremendous amount of computing resources.

The initial conditions (ICs) of cosmological simulations are constructed based on the Gaussian realizations that naturally introduce noise into the simulations. Although the introduced noises are physically motivated (i.e. due to inflation), we want to minimize them to provide a noiseless theoretical prediction to compare with observed data. Otherwise, we will waste the constraining power from the observation. A brutal way to reduce the noise would be generating simulations with larger volumes until the noise is well below the statistical error of the survey. However, this strategy will not be practical when the survey volume increases dramatically as DESI.
Some techniques have been developed to reduce the noise without running massive volume simulations. \cite{Angulo2016} proposed the fixed-amplitude technique to suppress the variance at large scales by modifying the ICs of simulations. In addition, it can further suppress sample variance using pairs of the fixed-amplitude simulations with initial phases differed by $\pi$ rad \citep{Pontzen2016}. The so-called paired-and-fixed method has been studied by a series of work \citep[e.g.][]{Villaescusa2018, Chuang2019, Klypin2020, Avila2022, Maion2022}.

Recently, the CARPool method, proposed by \cite{Chartier2021a}, takes a different approach. It reduces the noise by learning the calibrations from a large set of quasi-$N$-body simulations. While the fixed-amplitude method has only small improvement or no improvement in the precision at small scales, e.g. $k > 0.2\hMpc$, the CARPool method still has a significant gain at even smaller scales.

In this work, we apply the CARPool method to the \textsc{AbacusSummit} simulations, an extensive simulation suite generated on the Summit supercomputer\footnote{https://www.olcf.ornl.gov/summit/}. To do so, we prepare a set of \textsc{FastPM} simulations matching the configuration, including the ICs of \textsc{AbacusSummit} simulations. While the CARPool method has been validated for the statistics of dark matter particles \citep{Chartier2021a}, we focus on dark matter haloes that are expected to host DESI-like galaxy samples, e.g. ELGs \citep{Gonzalez2018, Avila2020} and LRGs \citep{Hernandez2021,Zhou2021}.
The \textsc{AbacusSummit} suite includes various cosmology models, but only the base cosmology has the largest volume which is $200\Gpchcube$ (i.e. 10 times DESI volume). In this work, we also extend the CARPool method to increase the effective volume of simulations other than the base cosmology as well.
Our work paves the way for providing the most precise and accurate galaxy clustering predictions based on $N$-body simulations for DESI or future surveys.

This paper is organized as follows. In Section~\ref{sec:simulation} we describe the simulations used in this study. We briefly describe the CARPool method in Section~\ref{sec:carpool}. We show the CARPool performance for the halo two-point and three-point clustering statistics in Section~\ref{sec:clustering}. We show the results of extending the method to different cosmologies in Section~\ref{sec:diff_cosmo}. Finally, in Section~\ref{sec:conclusion}, we present the conclusions and discussions.

\section{Simulations}\label{sec:simulation}
We describe the $N$-body simulations and quasi-$N$-body simulations we use or prepare for this study.
\subsection{\textsc{AbacusSummit} Simulations}
\textsc{AbacusSummit}\footnote{https://abacussummit.readthedocs.io/en/latest/} \citep{Maksimova2021} is a suite of high-fidelity $N$-body simulations based on the \textsc{Abacus} $N$-body code \citep{Metchnik2009, Garrison2016, Garrison2018, Garrison2019, Garrison2021}. They were generated on the Summit supercomputer at the Oak Ridge Leadership Computing Facility for the scientific goals of DESI survey. \textsc{AbacusSummit} consists of simulations that span different cosmologies, box sizes and mass resolutions. The base cosmology, denoted as c000, is the flat $\Lambda$ cold dark matter ($\Lambda$CDM) model constrained from Planck 2018 \citep{Planck2018}. In c000, there are 25 base boxes, each of which is a $2\Gpch$ box with $6912^3$ particles and particle mass resolution $2.11\times 10^9 \Msunh$.
There are four secondary cosmologies (c001--c004) and each cosmology has six base boxes that share the ICs of the first six boxes (ph000--ph005) in c000. We show the parameters of cosmologies c000, c002, and c004 in Table \ref{tab:cosmologies}.
\begin{table*}
    \caption{Parameters of the \textsc{AbacusSummit} cosmologies. c000 is the flat $\Lambda$CDM based on Planck 2018; c002 and c004 are two of the secondary cosmologies with some parameters different from c000's. c002 is a thawing dark energy model with $w_0=-0.7$ and $w_a=-0.5$. c004 has lower clustering amplitude, i.e. smaller $A_s$ and $\sigma_8$ than those of c000. More details of the \textsc{AbacusSummit} cosmological models can be found in \citet{Maksimova2021}.}
    \centering
    \begin{tabular}{|c|c|c|c|c|c|c|c|c|c|c|c|c|c|}
    \hline
        Cosmology & $\Omega_b h^2$ & $\Omega_{\text{cdm}} h^2$ & $h$ & $A_s$ & $n_s$ & $\alpha_s$ & $N_\text{ur}$ & $N_\text{ncdm}$ & $\Omega_\text{ncdm} h^2$ & $w_\text{0,\,fld}$ & $w_\text{a,\, fld}$ & $\sigma_{8_m}$ & $\sigma_{8_{cb}}$ \\
    \hline
        c000 & 0.02237 & 0.1200 & 0.6736 & 2.0830e-9 & 0.9649 & 0.0 & 2.0328 & 1 & 0.00064420 & -1.0 & 0.0 & 0.807952 & 0.811355 \\
    \hline
        c002 & 0.02237 & 0.1200 & 0.6278 & 2.3140e-9 &0.9649& 0.0 & 2.0328 & 1 & 0.00064420 & -0.7 & -0.5 & 0.808189 & 0.811577 \\
    \hline
        c004 & 0.02237 & 0.1200 & 0.6736 & 1.7949e-9 & 0.9649 & 0.0 & 2.0328 & 1 & 0.00064420 & -1.0 & 0.0 & 0.749999 & 0.753159 \\
    \hline    
    \end{tabular}
    
    \label{tab:cosmologies}
\end{table*}

\textsc{AbacusSummit} uses the highly efficient on-the-fly \textit{Competitive Assignment to Spherical Overdensities} (\textsc{CompaSO}) group finder \citep{Hadzhiyska2021} and outputs halo catalogues at 12 primary redshifts, $z=3.0,\, 2.5,\, 2.0,\, 1.7,\, 1.4,\, 1.1,\, 0.8,\, 0.5,\, 0.4,\, 0.3,\, 0.2$, and $0.1$, as well as at 21 secondary redshifts. Due to the large amount of data, we only focus on the halo catalogues at $z=1.1$ which is the median redshift where the primary targets (ELGs) of DESI will be observed ($z=0.6$--$1.6$). We do not use the "cleaned" version of the \textsc{CompaSO} catalogues, which are not available when we started this work. The details of the cleaning method are described in \cite{Bose2021,Hadzhiyska2021}. In summary, based on halo merger trees, the cleaning method can remove unphysical haloes identified by \textsc{CompaSO}. The misidentification is mainly caused by two sides. One is due to the halo dynamical processes such as fly-bys, partial mergers, and splits, and the other is from the strict spherical overdensity criterion that can overly deblend single haloes into two or more components. After cleaning, the number of haloes will be decreased by a few per cent and mainly for low-mass haloes ($M_{\text{halo}}\sim 10^{11}\Msunh$). We believe that whether using the cleaned or uncleaned version of the halo catalogues should not affect our main conclusions, though the cleaned version will cause some difference on halo correlation function or power spectrum around the scale of one-halo to two-halo transition and small difference on the overall halo bias. For example, the halo bias from the cleaned catalogue with $M_{\text{halo}}\sim 10^{11.5}\Msunh$ will be $\sim 6$ per cent lower \citep[fig. 10 in][]{Bose2021} compared with that of the uncleaned one. In this work, we analyse the \textsc{AbacusSummit} halo statistics with the assistance of the package \textsc{abacusutils}\footnote{https://abacusutils.readthedocs.io/en/latest/}.

\subsection{DESI-\textsc{FastPM} Simulation}
We choose \textsc{FastPM}\footnote{https://github.com/fastpm/fastpm} \citep{Feng2016} as the surrogate to pair with \textsc{AbacusSummit}. \textsc{FastPM} is a fast simulation method to approximate clustering from $N$-body solvers. It implements the particle mesh (PM) scheme \citep{Quinn1997} with modified kick and drift factors to guarantee the accuracy of the linear displacement at large scales. The accuracy of \textsc{FastPM} is mainly determined by the particle mass resolution $m_0$, the initial redshift $z_0$, the number of time-steps $T$, and the force resolution which is parametrized as the ratio of the force mesh size over the number of particles along one axis of the simulation box, denoted as B. 

\subsubsection{Configuration of DESI-\textsc{FastPM} Simulations}
Apart from the accuracy, we need to consider the computational cost since a large number of \textsc{FastPM} simulations are required to construct covariance matrices and to do cosmological analysis. 
Therefore, we first need to determine the configuration parameters to balance its accuracy and computational cost.
In order to pair \textsc{FastPM} with \textsc{AbacusSummit}, we set the \textsc{FastPM} box size the same as the \textsc{AbacusSummit} base runs, i.e. $\Lbox=2\Gpch$.
Given a \textsc{FastPM} simulation with some certain configuration parameters, we compare its matter power spectrum with that of \textsc{AbacusSummit} at redshift 0.2.
We set the number of particles in \textsc{FastPM} as $N_p=5184$ per side; hence, the particle mass is $m_0=5\times 10^9\Msunh$, about 2.4 times larger than that of \textsc{AbacusSummit} for the base cosmology.

While the default version of \textsc{FastPM} uses an extra particle species to simulate massive neutrinos \citep{Bayer2020}, which are labelled as ncdm (not-cold dark matter), we use a modified version to treat ncdm the same as that in \textsc{AbacusSummit}, i.e. the effect of massive neutrinos only contributes to the Hubble expansion rate but not to the gravitational forces from clustering. 

In the end, we find a reasonable set of configuration parameters for the massive production of \textsc{FastPM} simulations.
We run \textsc{FastPM} from the initial redshift $z_0=19$ with the second-order Lagrangian perturbation theory (2LPT) IC to the final redshift $z=0.1$, with $40$ time-steps linearly separated in scale $a$. We set the PM size parameter $B=2$.
We validate such choice of the configuration parameters in Appendix \ref{sec:appendix_fastpm_parameters}.

\subsubsection{ICs and Cosmologies}\label{sec:fastpmIC}
Using the \textsc{AbacusSummit} base cosmology (c000), we have produced a bunch of \textsc{FastPM} simulations including
\begin{itemize}
    \item 25 boxes with the \textsc{AbacusSummit} ICs; 
    \item 201 boxes with independent ICs;
    \item 237 boxes with the fixed-amplitude ICs.
\end{itemize}
The technique of fixed amplitude of the initial density field \citep{Angulo2016} is a method to effectively suppress the sample variance, and hence the number of realizations is greatly reduced to reach a certain precision at large scales. Different from the Rayleigh distributed amplitude in the Gaussian density field, the amplitude of the density fluctuation is fixed to be the square root of the input linear power spectrum, i.e.
\begin{align}
    \delta^L(\textit{\textbf{k}}) = \sqrt{P(k)}\exp(i\theta(\textit{\textbf{k}})),
\end{align}
and $\theta(\textit{\textbf{k}})$ is the phase uniformly distributed in $(0, 2\pi]$. 
There have been multiple studies \citep[e.g.][]{Villaescusa2018, Chuang2019} showing the unbiasedness of several statistics from the fixed-amplitude method, compared with the results simulated from the Gaussian initial density field. 

In terms of \textsc{AbacusSummit} secondary cosmologies, we have produced
\begin{itemize}
\item 25 boxes for both c002 and c004 (50 boxes in total) with the same white noises as the base cosmology c000.
\end{itemize}

\subsubsection{Running DESI-\textsc{FastPM} on NERSC}
As a project in the DESI collaboration, we run all the \textsc{FastPM} simulations on Cori supercomputer at the National Energy Research Scientific Computing Center (NERSC)\footnote{https://www.nersc.gov/}. NERSC is one of the largest facilities in the world devoted to basic scientific research. We use KNL computing nodes for the computation. Each KNL node has 68 CPU cores and 96 GB memory. We assign 1152 nodes and 36 MPI tasks per node for each \textsc{FastPM} simulation. Each simulation takes about 50 min of wall-clock time. We down-sample dark matter particles and compress particle information into integers. Such process uses the same amount of nodes and about 7 min of wall-clock time. The total cost of the simulations is about 24 million NERSC hours. For each simulation, we utilize about $55000$ GB temporary space from the Burst Buffer\footnote{It uses flash or SSD (solid-state drive) array to achieve high speed on I/O.} to store the output. Once the simulation is finished, the output is transported from the Burst Buffer to the disk automatically in the backend.

\subsubsection{Products and Storage}

We output 12 snapshots of \textsc{FastPM} dark matter catalogues and halo catalogues at the same primary redshifts of \textsc{AbacusSummit}. Finding haloes from the dark matter field, we use the \textsc{FastPM} internal Friends-of-Friends algorithm with the linking length equal to $0.2$ times of the mean separation of particles. We store haloes with mass larger than $M_{\mathrm{halo}}=5\times 10^{10}\Msunh$. Since we need to run hundreds of simulations, the total output data will take too much storage. To save disk space, we down-sample dark matter particles by $1/27$ and store their positions and velocities in 1-byte integer for future usage, e.g. matter density field and weak lensing light-cone construction. Specifically, the position is stored in the form of displacement from Lagrangian lattice. The float displacement and velocity are converted into bits by the error function $\mathrm{erf}(sx)$. We minimize the loss in the conversion by choosing the optimal scaling factor $s$ in the error function for each redshift. We store information of halo catalogues, including positions, velocities, masses, inertial tensors, velocity dispersion and angular momenta, using 4-byte floating points. For each simulation, the total data size is about 810 GB with 350 GB for dark matter particles and 460 GB for haloes.

\section{CARPool Method}\label{sec:carpool} 

Based on the principle of control variates \citep{Rubinstein1985, Avramidis1993, deOPortaNova1993}, \citet{Chartier2021a} applied it to construct variance-reduced observables of large-scale structure clustering based on simulations. The approach is named as \textit{Convergence Acceleration by Regression and Pooling}, short for CARPool. It pairs a few $N$-body simulations and surrogates which approximate $N$-body simulations and share the ICs from the $N$-body simulations. The method of CARPool can be summarized by the equation as 
\begin{align}
    x = y - \beta(c-\mu_c),\label{eq:carpool}   
\end{align}
where $x$ is constructed as a representative of $y$ which is some observable from an $N$-body simulation, e.g. \textsc{AbacusSummit} in our case, $c$ is the same observable from the paired surrogate, e.g. \textsc{FastPM}, $\beta$ is the control variate, and $\mu_c$ is the mean of $c$. If $\mu_c$ is unknown, we can estimate it from a separate set of surrogates that do not share the ICs from the $N$-body simulations, i.e.
\begin{align}
    \hat{\mu}_c = \frac{1}{M}\sum_{j=1}^M c_j,
\end{align}
where $M$ is the number of surrogates. In our case, we have 201 independent normal boxes and 237 boxes using the fixed-amplitude ICs (see Sec. \ref{sec:fastpmIC}), both of which can be used to estimate $\mu_c$.
The computational cost of CARPool can be much cheaper than the traditional method which usually needs to run a large number of $N$-body simulations.

First, by design, $x$ is unbiased relative to $y$, as the ensemble average of equation (\ref{eq:carpool}) gives the expectation of $x$ equal to that of $y$, i.e. $\overline{x}=\overline{y}$. In the following we use overbars to denote means. Secondly, we can find the best $\beta$ to minimize the variance of $x$. If y is a vector, e.g. halo power spectrum, we have
\begin{align}
    \beta^{\star} = \Sigma_{yc} \Sigma_{cc}^{-1},
\end{align}
where $\Sigma_{yc}$ is the cross-covariance matrix between $y$ and $c$, and $\Sigma_{cc}$ is the covariance matrix of $c$, i.e.
\begin{align}
    \Sigma_{yc} &= \frac{1}{N-1}\sum_{i=1}^N (y_i-\overline{y})(c_i - \overline{c})^T,\\
    \Sigma_{cc} &= \frac{1}{N-1}\sum_{i=1}^N (c_i - \overline{c})(c_i - \overline{c})^T,
\end{align}
where $N$ is the number of paired simulations, and $\overline{y}$ and $\overline{c}$ denote the mean of the paired $N$-body simulations and surrogates, respectively.

As suggested by \cite{Chartier2021a}, if there are not many paired simulations, which is our case, it is better to choose the diagonal form of $\beta^{\star}$ to give better performance of CARPool, i.e.
\begin{align}\label{eq:beta_diag}
    \beta^{\text{diag}} = \text{Diag}(\beta^{\star})=\frac{\sigma_{yc}^2}{\sigma_c^2},
\end{align}
where $\sigma_{yc}^2$ is the cross-correlation between $y$ and $c$ from the same bins, and $\sigma^2_c$ is the variance of $c$. Our following results are all based on the diagonal form of $\beta$; hence, we ignore the superscript of $\beta^{\text{diag}}$ hereafter. 

The variance of the calibrated variable $x$ is
\begin{align}
    \sigma_{x}^2=\sigma_y^2-2\beta\sigma_{yc}^2+\beta^2\sigma_c^2+\beta^2\sigma_{\hat{\mu}_c}^2,\label{eq:sigma_x}
\end{align}
where we account for the variance of $\hat{\mu}_c$ estimated from surrogates,
\begin{align}
    \sigma_{\hat{\mu}_c}^2=\frac{1}{M(M-1)}\sum_{j=1}^M (c_j-\hat{\mu}_c)^2.
\end{align}
Note that there is no cross-correlation between $y$ and $\hat{\mu}_c$ in equation (\ref{eq:sigma_x}), since they are generated from different ICs.
Once substituting $\beta$ in equation (\ref{eq:sigma_x}), we obtain
\begin{align}
    \sigma_{x}^2=\left(1-\rho^2_{yc}\right)\sigma_y^2+\frac{\sigma_{yc}^4}{\sigma_c^4}\sigma_{\hat{\mu}_c}^2,
\end{align}
where $\rho_{yc}$ is the Pearson correlation coefficient between y and c, i.e. $\rho_{yc}=\frac{\sigma_{yc}^2}{\sigma_y\sigma_c}$. 
One can see that $\sigma_x$ could be very small if $y$ and $c$ are highly correlated ($\rho_{yc}$ close to 1.0), and if $\sigma_{\hat{\mu}_c}$ is small too.
In addition, we can derive the variance of $\overline{x}$ by
\begin{align}
    \sigma_{\overline{x}}^2=\frac{1}{N}\left(1-\rho^2_{yc}\right)\sigma_y^2+\frac{\sigma_{yc}^4}{\sigma_c^4}\sigma_{\hat{\mu}_c}^2. \label{eq:sigma_carpool_mean}
\end{align}
where the scaling factor $1/N$ should not be applied to the term with $\sigma_{\hat{\mu}_c}^2$. This is because we use the same set of surrogates to estimate $\hat{\mu}_c$ when we calculate $x$ for each paired simulation. Taking the mean of the CARPool result $x$ can only suppress the sample variances of the $N$-body simulations, but not that of the surrogate mean $\hat{\mu}_c$.

One can also derive the effective volume by comparing $\sigma_{\overline{x}}$ and $\sigma_{\overline{y}}$, 
where $\sigma_{\overline{y}}^2=\sigma_y^2/N$ and $N=25$ in this study. Note that the combination of 25 \textsc{AbacusSummit} base simulations has a total volume of $200\Gpchcube$ which is about 10 times the effective volume of DESI survey. Thus, the effective volume gained from CARPool can be given by
\begin{align}
    V_{\text{eff}}&=\frac{\sigma_{\overline{y}}^2}{\sigma_{\overline{x}}^2} 200\Gpchcube =\frac{\sigma_{\overline{y}}^2}{\sigma_{\overline{x}}^2}10\ V_{\textrm{DESI}}\label{eq:veff}
\end{align}

\section{Application on Halo Clustering}\label{sec:clustering}
\begin{figure*}
    \centering
    \includegraphics[width=1.0\linewidth]{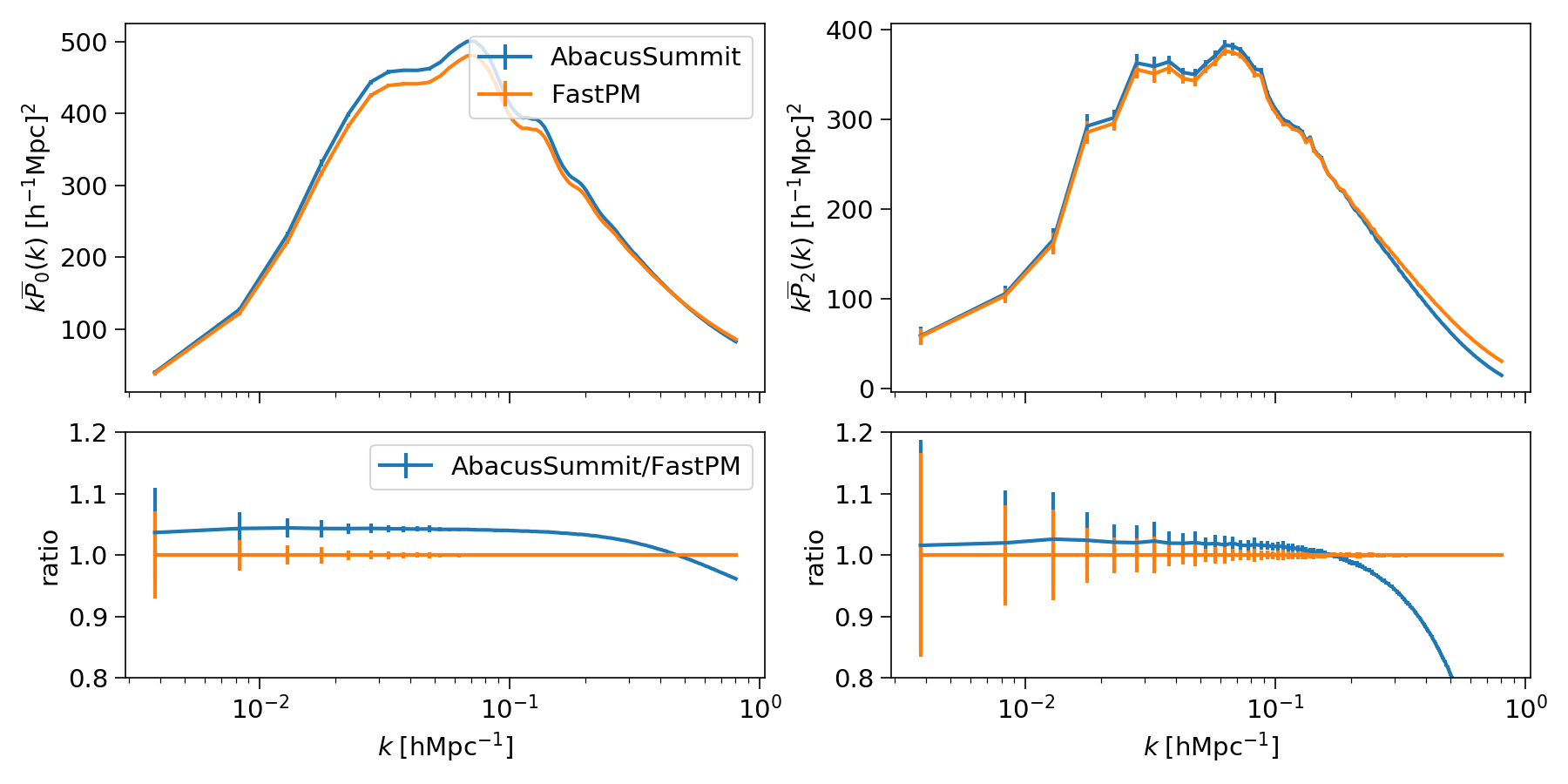}
    \caption{\textit{Upper panels}: the mean halo power spectrum monopole (left-hand panel) and quadrupole (right-hand panel) from 25 halo catalogues of the paired \textsc{AbacusSummit} and \textsc{FastPM} simulations. The halo catalogues are at redshift 1.1 and contain haloes with mass larger than $10^{11}\Msunh$. The blue lines are for \textsc{AbacusSummit} and the orange lines are for \textsc{FastPM}. \textit{Lower panels}: the ratio of the mean between \textsc{AbacusSummit} and \textsc{FastPM}. The error bars represent the standard deviations of the mean multipoles.}
    \label{fig:pk_mcut1.e11}
\end{figure*}

\cite{Chartier2021a} have applied the CARPool method on the clustering statistics of dark matter in real space. In our study, we extend their analysis to halo clustering with two-point statistics in redshift space and three-point statistics in real space. 

\subsection{Halo power spectrum}
We study halo catalogues at redshift $z=1.1$ which is a typical redshift since DESI will observe 10 million ELGs at redshift $0.6 < z < 1.6$. We apply a mass cut and select haloes with mass larger than $10^{11}\Msunh$ which is the expected minimum halo mass hosting ELGs. Apart from the differences on the dark matter simulations, the halo finders from \textsc{AbacusSummit} and \textsc{FastPM} are different as well. Therefore, we do expect some difference on the halo clustering even on large scales. We first study the halo power spectrum $P(\bmath{k})$ defined as
\begin{align}
    \langle \delta(\bmath{k})\delta(\bmath{k^{\prime}})\rangle \equiv (2\pi)^3\delta_D(\bmath{k} + \bmath{k^{\prime}})P(\bmath{k}),
\end{align}
where $\delta(\bmath{k})$ is the halo number density fluctuation as a function of wavevector $\bmath{k}$ in Fourier space and $\delta_D$ is the Dirac delta function. In redshift space, power spectrum is not isotropic due to the peculiar velocity along the line of sight, hence, it can be decomposed into multipoles
\begin{align}
    P_{\ell}(k) = (2\ell + 1)\int_0^1 P(k, \mu) L_{\ell}(\mu) d\mu, \label{eq:pk_ell}
\end{align}
where $L_{\ell}(\mu)$ is the Legendre polynomial of order $\ell$, and $\mu$ is the cosine of the angle between $\bmath{k}$ and the line of sight, i.e. 
\begin{align}
   \mu = k_{\|}/k,\;\; k=\sqrt{k_{\perp}^2 + k_{\|}^2},
\end{align}
with $k_{\perp}$ and $k_{\|}$ being the components of $\bmath{k}$ perpendicular and parallel to the line of sight, respectively. In our study, we present the results of monopole and quadrupole that are widely analysed in galaxy surveys. 

We calculate the halo power spectrum multipoles via \textsc{nbodykit}\footnote{https://nbodykit.readthedocs.io} \citep{Hand2018}. For the calculation, we paint haloes in a mesh with $1024^3$ cells using the triangular-shaped cloud mass assignment window. We eliminate the aliasing effect with the interlacing technique which can compensate the window function effect. We set $120$ linear $\mu$ bins in the range $[0,\,1]$ and integrate the anisotropic power spectrum over $\mu$ (based on equation \ref{eq:pk_ell}) to obtain the power spectrum multipoles. We study the multipoles in the $k$ range from $0.0038$ to $0.8025\hMpc$ which is half of the Nyquist frequency. The total number of $k$ bins is 161 with the interval $0.005\hMpc$. To be compact, for the following, we show only the CARPool results from the power spectrum monopole ($\ell=0$) and quadrupole ($\ell=2$) while ignoring the hexadecapole which has larger statistical noise.

Fig. \ref{fig:pk_mcut1.e11} shows the mean halo power spectrum multipoles calculated from 25 pairs of \textsc{AbacusSummit} and \textsc{FastPM} catalogues, respectively. In the upper panels, we compare the overall shapes of monopoles (left-hand panel) and quadrupoles (right-hand panel) with the standard deviations of the mean from the paired simulations. In the lower panels, we show the ratio of the mean from \textsc{AbacusSummit} and \textsc{FastPM}. For the monopole, there is a constant bias about $4.2\%$ at large scales. Although there is noticeable difference on the halo number densities from the two simulations\footnote{With mass cut $10^{11}\Msunh$, \textsc{AbacusSummit} has halo number density $0.036\hMpccube$, $36\%$ higher than that of \textsc{FastPM}.}, we have checked that using abundance matching can only reduce the bias to $3.7\%$. 
Instead, if we use the cleaned \textsc{AbacusSummit} halo catalogues, the bias decreases to $1.2\%$. 
Thus, we argue that the constant bias at large scales is mainly due to different halo finders in the two simulations.
As a supplement, we show the power spectrum of the cleaned haloes in Appendix \ref{appendix:mc_am_hc}. At small scales, the power spectrum of \textsc{AbacusSummit} is smaller than that of \textsc{FastPM}. This is caused by the underestimation of the small-scale damping of redshift-space distortions (RSDs; i.e. the Fingers-of-God effect) in \textsc{FastPM}, since it is not able to trace the velocity field of particles precisely at small scales.

\begin{figure}
    \centering
    \includegraphics[width=1\linewidth]{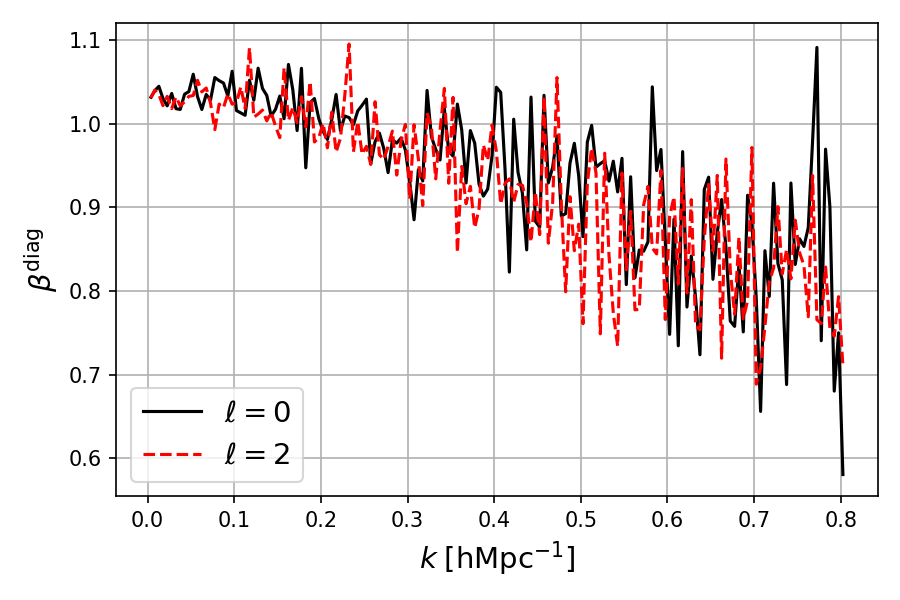}
    \caption{$\beta^{\text{diag}}$ of the halo power spectrum multipoles, i.e. monopole $\ell=0$ and quadrupole $\ell=2$, from the paired \textsc{AbacusSummit} and \textsc{FastPM} halo catalogues with mass cutoff $10^{11}\Msunh$. The cross-correlation between the multipoles of \textsc{AbacusSummit} and \textsc{FastPM} is high on large scales, as $\beta^{\text{diag}}$ is close to 1.0, and it decreases as the scale becomes smaller. The shape of $\beta^{\text{diag}}$ is similar between the monopole and quadrupole.}
    \label{fig:beta_pk_mcut1e11}
\end{figure}

\begin{figure*}
    \centering
    \includegraphics[width=1\linewidth]{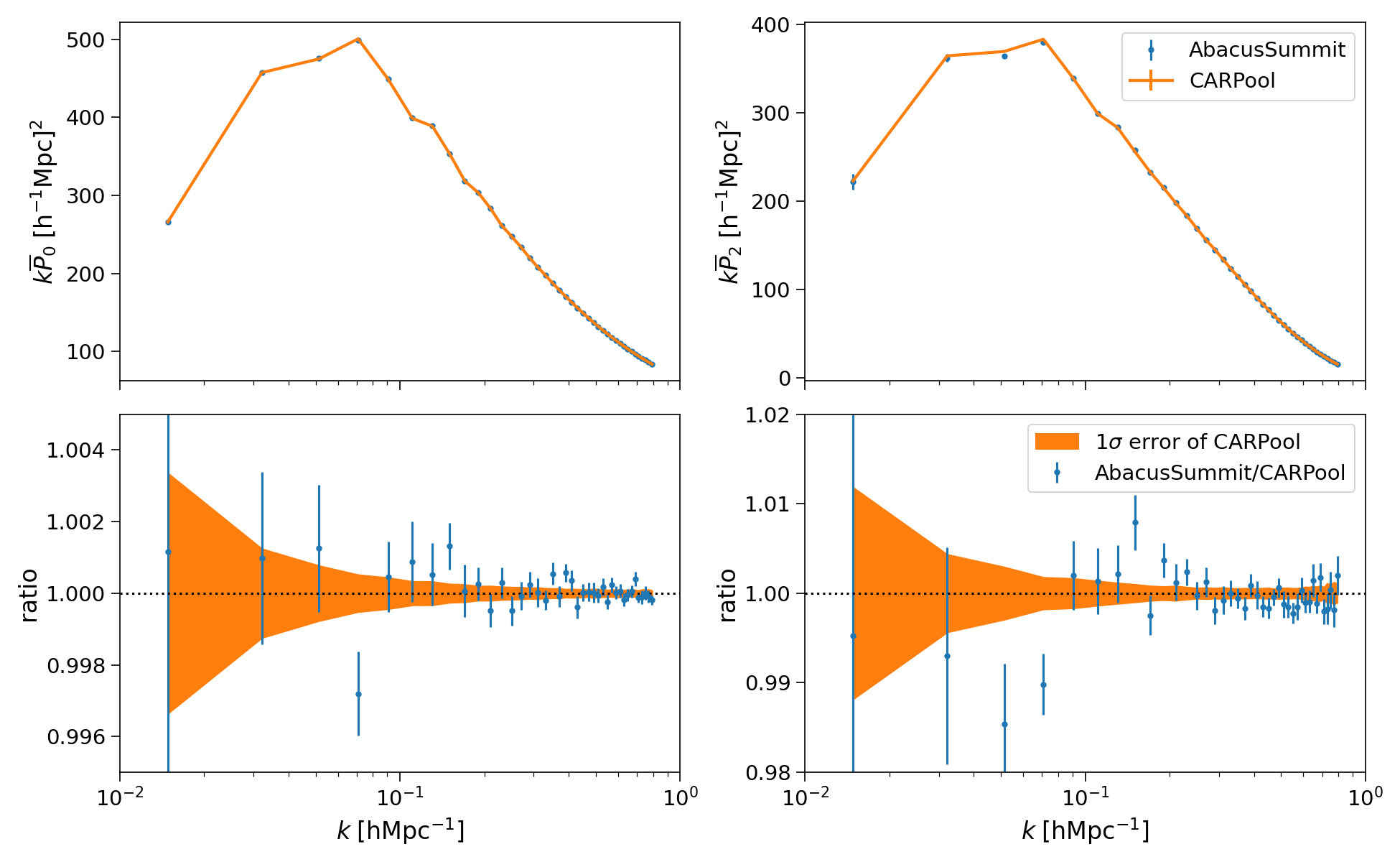}
    \caption{\textit{Upper panels}: the mean of the halo power spectrum monopole (left-hand panel) and quadrupole (right-hand panel) from 25 \textsc{AbacusSummit} simulations compared with the mean of $P_x$ from CARPool. Error bars have been scaled to represent the standard deviation of the mean. \textit{Lower panels}: the ratio of the mean from \textsc{AbacusSummit} and CARPool, shown as the blue points, compared with the $1\sigma$ error of the mean $P_x$, shown as the orange shades. There is no bias on the mean power spectrum multipoles from CARPool, and the sample variance is reduced significantly.}
    \label{fig:px_mcut1e11}
\end{figure*}

\subsubsection{Performance of CARPool method}
Based on the power spectrum multipoles from the paired \textsc{AbacusSummit} and \textsc{FastPM} halo catalogues, we can calculate their cross-correlation, as well as the variance from each simulation. From equation (\ref{eq:beta_diag}), we obtain $\beta^{\text{diag}}$ for the monopole and quadrupole, as shown in Fig. \ref{fig:beta_pk_mcut1e11}. We find that the overall shapes of $\beta^{\text{diag}}$ for the monopole and quadrupole are similar, i.e. it is close to $1.0$ on large scales (small $k$), which is due to the high cross-correlation between the multipoles from the paired \textsc{AbacusSummit} and \textsc{FastPM} simulations. This is guaranteed since the paired simulations share the same ICs and have similar information of the large scale structures. As the scale becomes smaller, the cross-correlation decreases, due to the differences from the non-linear evolution, the halo finders, etc., between the two simulations. The noises in the curves are due to the fact that we are using only 25 pairs of simulations to determine the variances.

We compare the influence on $\beta^{\text{diag}}$ from mass cut, abundance matching, as well as the halo cleaning of \textsc{AbacusSummit} in Appendix \ref{appendix:mc_am_hc}. There is negligible improvement from abundance matching, whereas using the cleaned \textsc{AbacusSummit} catalogues improves $\beta^{\text{diag}}$ closer to 1.0 by a few per cent in the range $k<0.2\hMpc$. We believe that neither using abundance matching nor cleaned haloes of \textsc{AbacusSummit} will influence our results much.

For given paired power spectrum multipoles from \textsc{AbacusSummit} and \textsc{FastPM}, we construct the new power spectrum multipoles $P_{x,\,\ell}$ via CARPool,
\begin{align}
    P_{x,\,\ell} = P_{\textsc{AbacusSummit},\,\ell} - \beta (P_{\textsc{FastPM},\, \ell} - \overline{P}^{\prime}_{\textsc{FastPM},\,\ell}), \label{eq:px_abacus}
\end{align}
where $\overline{P}^{\prime}_{\textsc{FastPM},\,\ell}$ with an overline denotes the mean halo power spectrum multipoles calculated from a separate set of \textsc{FastPM} simulations, e.g. 201 \textsc{FastPM} simulations with random ICs. As discussed in Section \ref{sec:carpool}, the mean $P_{x,\,\ell}$ should be unbiased compared with the mean $P_{\textsc{AbacusSummit},\,\ell}$, and the statistical error of $P_{x,\,\ell}$ should be smaller than that of $P_{\textsc{AbacusSummit},\,\ell}$. We check these by calculating the mean of $P_{x,\,\ell}$ over 25 realizations and the standard deviation, 
\begin{align}
    \overline{P}_{x,\,\ell} &= \frac{1}{N} \sum_i^N P^{i}_{x,\,\ell}, \label{eq:px_mean}\\
    \sigma_{P_{x,\, \ell}}^2 &= \frac{1}{N-1} \sum_i^N (P_{x,\, \ell}^i - \overline{P}_{x,\,\ell})^2. \label{eq:sudo_sigma_px}
\end{align}
Since we estimate the mean of \textsc{FastPM} multipoles from a limited number of realizations, the standard deviation of $P_{x,\,\ell}$ (equation \ref{eq:sudo_sigma_px}) is underestimated. Considering the error of $\overline{P}_{\textsc{FastPM},\,\ell}^{\prime}$, we have
\begin{align}
    \sigma_{P^{\prime}_{x, \,\ell}}^2 = \sigma_{P_{x,\,\ell}}^2 + \beta^2 \sigma^2_{\overline{P}^{\prime}_{\textsc{FastPM}, \,\ell}}.\label{eq:sigma_px}
\end{align}
And for the standard deviation of the mean of $P_{x,\,\ell}^{\prime}$, it is
\begin{align}
    \sigma_{\overline{P}^{\prime}_{x,\,\ell}}^2 = \frac{1}{N}\sigma_{P_{x,\,\ell}}^2 + \beta^2 \sigma^2_{\overline{P}^{\prime}_{\textsc{FastPM}, \,\ell}},\label{eq:sigma_px_mean}
\end{align}
From now on, we ignore the subscript $\ell$ of multipoles for simplicity.

In the upper panels of Fig. \ref{fig:px_mcut1e11}, we show the mean multipoles with the standard deviations from CARPool as the orange lines. To avoid crowdedness of data points at small scales, we redo $k$ binning for the power spectrum multipoles using a larger $k$ interval. The left-hand panel is for the monopole and the right-hand panel is for the quadrupole. In the standard deviation of the CARPool mean, we take account of the error of the \textsc{FastPM} mean, which is calculated from 201 sets of regular \textsc{FastPM} simulations. We compare the results with the mean multipoles from 25 \textsc{AbacusSummit} halo catalogues shown as the blue points. We see that they agree well with each other in $1\sigma$ error over all the scales except for some points at large scales, which is just due to cosmic variance. In the lower panels, we show the ratio of the mean between \textsc{AbacusSummit} and CARPool, shown as the blue points. The orange shaded region denotes the noise-to-signal ratio, i.e. $\sigma_{\overline{P}_x}/{\overline{P}_x}$ from CARPool. The blue points fluctuate near unity over all the scales, illustrating the unbiasedness of $P_x$. In addition, the standard deviation of $\overline{P}_x$ from CARPool is much smaller than the original standard deviation of $\overline{P}_{\textsc{AbacusSummit}}$. 

\begin{figure*}
    \centering
    \includegraphics[width=1\linewidth]{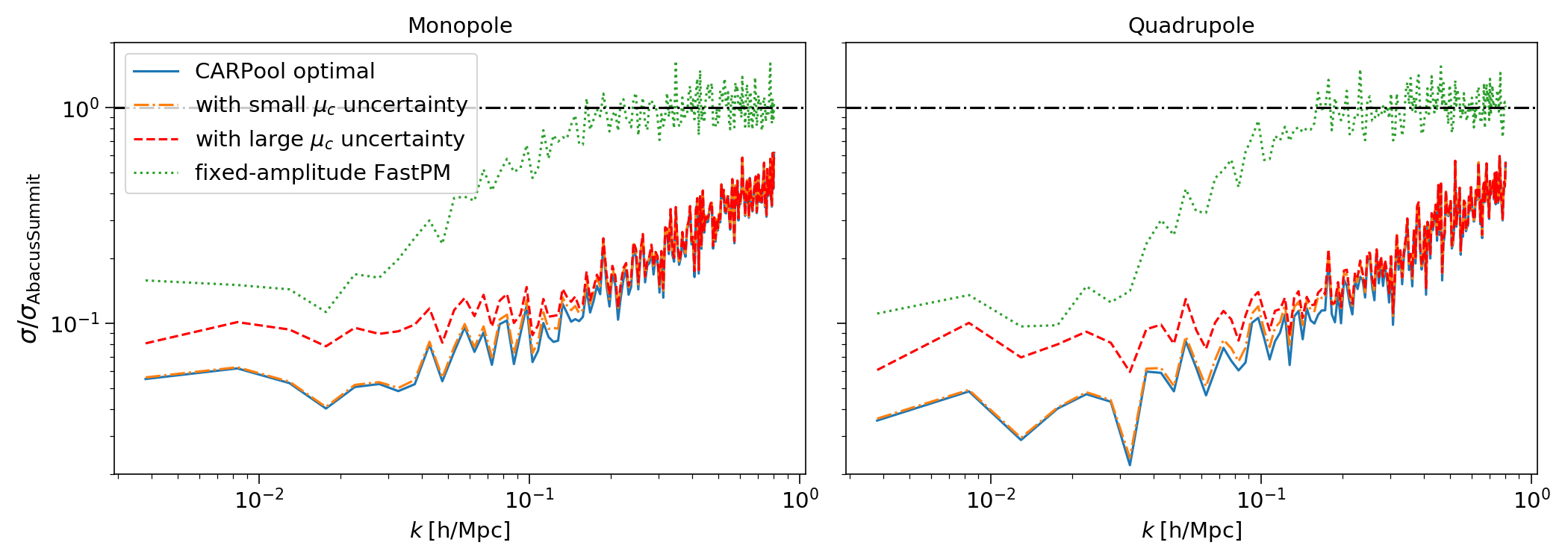}
    \caption{The reduction of the sample variance on the halo power spectrum multipoles from CARPool and the method of the fixed-amplitude IC. The left-hand panel is for the monopole and the right-hand panel is for the quadrupole. In each panel, the blue solid line shows the optimal gain expected from CARPool if the mean of the \textsc{FastPM} multipole in  (\ref{eq:px_abacus}) is known. The red dashed line is one realistic case considering the sample variance of the surrogate mean estimated from 201 nonfixed-amplitude \textsc{FastPM} realizations. The fixed-amplitude method can also effectively reduce the sample variance, especially on large scales. We calculate the standard deviation of 237 fixed-amplitude \textsc{FastPM} realizations and compare it with that of the non-fixed-amplitude ones, shown as the green dotted lines. Using CARPool with the surrogate mean from the fixed-amplitude \textsc{FastPM}, we can further increase the error reduction from the red dashed lines to the orange dot--dashed lines which almost overlap with the blue lines. The performance of CARPool on the monopole and quadrupole is similar, which is expected since the cross-correlation between \textsc{AbacusSummit} and \textsc{FastPM} is similar for the monopole and quadrupole, shown as $\beta^{\mathrm{diag}}$ in Fig. \ref{fig:beta_pk_mcut1e11}.}
    \label{fig:sigma_pkx_mcut1e11}
\end{figure*}

To quantify the reduction on the sample variance from CARPool, we compute the ratio of the standard deviation of $P_x$ over that of $P_{\textsc{AbacusSummit}}$, and show the results in Fig. \ref{fig:sigma_pkx_mcut1e11}. The left-hand panel is for the monopole and the right-hand panel is for the quadrupole. In each panel, the blue line is the optimal case where we assume that the surrogate mean, i.e. $\overline{P}^{\prime}_{\textsc{FastPM}}$, is known. The red dashed line is a realistic case once we consider the standard deviation of $\overline{P}^{\prime}_{\textsc{FastPM}}$ from 201 regular \textsc{FastPM} simulations. On large scales, the reduction of statistical error from CARPool is very significant, even after including the error of the surrogate mean. In our conservative case (red dashed line), CARPool can reduce the standard deviation to be 10 times smaller, for both monopole and quadrupole on large scales. As the scale goes smaller ($k>0.1\hMpc$), the gain from CARPool gradually decreases; however, it still has $50\%$ error suppression up to $k=0.8\hMpc$. 

Since the fixed-amplitude technique can also reduce statistical errors especially on large scales, it would be interesting to compare the performance with CARPool. In our project, we have run 237 realizations of \textsc{FastPM} with the fixed-amplitude ICs. We compare the standard deviation of the fixed-amplitude \textsc{FastPM} simulations with that of the \textsc{FastPM} simulations using the \textsc{AbacusSummit} ICs. The results are shown as the green dotted lines in Fig. \ref{fig:sigma_pkx_mcut1e11}. For the power spectrum multipoles, the error reduction from the fixed-amplitude method has a similar trend as the CARPool result, but is less significant than that of CARPool, and it quickly reduces to zero on smaller scales, in our case, $k>0.2\hMpc$. 
Because the error reduction from CARPool is limited by the error of surrogate mean, we confirm that if we replace the surrogate mean by the mean from the fixed-amplitude \textsc{FastPM} simulations, we can further reduce the error of the CARPool result. As a result, we obtain the gain close to the optimal case, shown as the orange dot--dashed lines, which nearly overlap with the blue lines. 

\begin{figure*}
    \centering
    \includegraphics[width=1\linewidth]{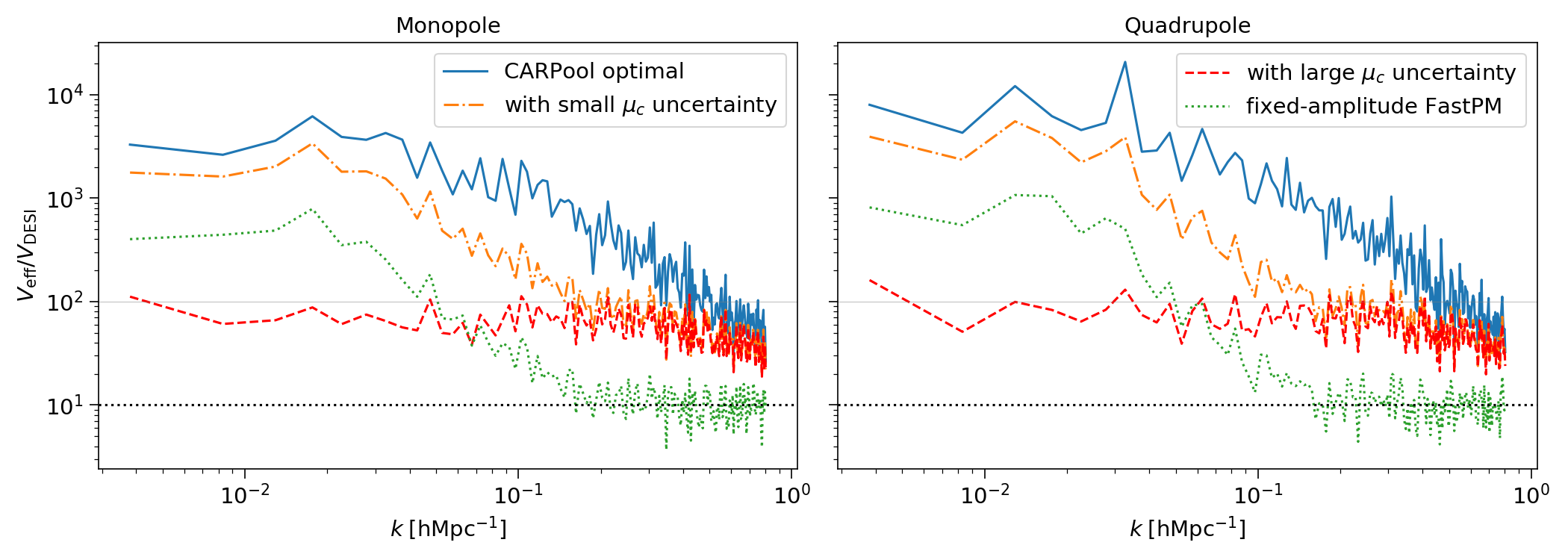}
    \caption{The increase of the effective volume from the combination of $25$ \textsc{AbacusSummit} simulations with CARPool compared with the result from the fixed-amplitude method. The left-hand (right-hand) panel is for the monopole (quadrupole). In each panel, the blue line represents the optimal case assuming no sample variance from the surrogate mean. The red dashed line takes account of the sample variance of the surrogate mean based on the non-fixed-amplitude \textsc{FastPM} catalogues. The orange dot--dashed line is the case if we reduce the sample variance of the surrogate mean using the fixed-amplitude \textsc{FastPM} catalogues. For comparison, we also show the effective volume from 25 fixed-amplitude \textsc{FastPM} catalogues as the green dotted line.}
    \label{fig:effective_volume}
\end{figure*}

The reduction on the sample variance corresponds to the increase of the effective volume. Based on equation (\ref{eq:veff}), we estimate the effective volume from the combination of 25 \textsc{AbacusSummit} boxes with CARPool and compare it with the total DESI 5 yr survey volume. We summarize our results in Fig. \ref{fig:effective_volume} with the left-hand panel for the monopole and the right-hand panel for the quadrupole. In each panel, the blue line is the optimal case from CARPool assuming that the surrogate mean is known. The red dashed line is a realistic case considering the error of $\overline{P}^{\prime}_{\textsc{FastPM}}$ from the non-fixed-amplitude \textsc{FastPM}, where we can increase the effective volume about $100$ times over all the $k$ range. Different from the optimal case that has a larger increase of the effective volume on larger scales, in the realistic case the error of the surrogate mean suppresses such increase as the error of the surrogate mean dominates in equation (\ref{eq:sigma_px_mean}). The green dotted line represents the effective volume from 25 fixed-amplitude \textsc{FastPM} boxes. On large scales, it is higher than the red dashed line but drops quickly as the scale goes smaller and turns to no gain on scales of $k$ larger than $0.2\hMpc$. If we replace the surrogate mean from the non-fixed-amplitude \textsc{FastPM} by that of the fixed-amplitude ones, the error of the surrogate mean is dramatically suppressed. We obtain the result shown as the orange dot--dashed line which can be understood as the combination of the red and green lines. With CARPool and the fixed-amplitude \textsc{FastPM} simulations, we can extend the effective volume of 25 \textsc{AbacusSummit} simulations to $10^2 \sim 10^3$ times the DESI volume with the dependence on scales.    

\begin{figure*}
    \centering
    \includegraphics[width=1\linewidth]{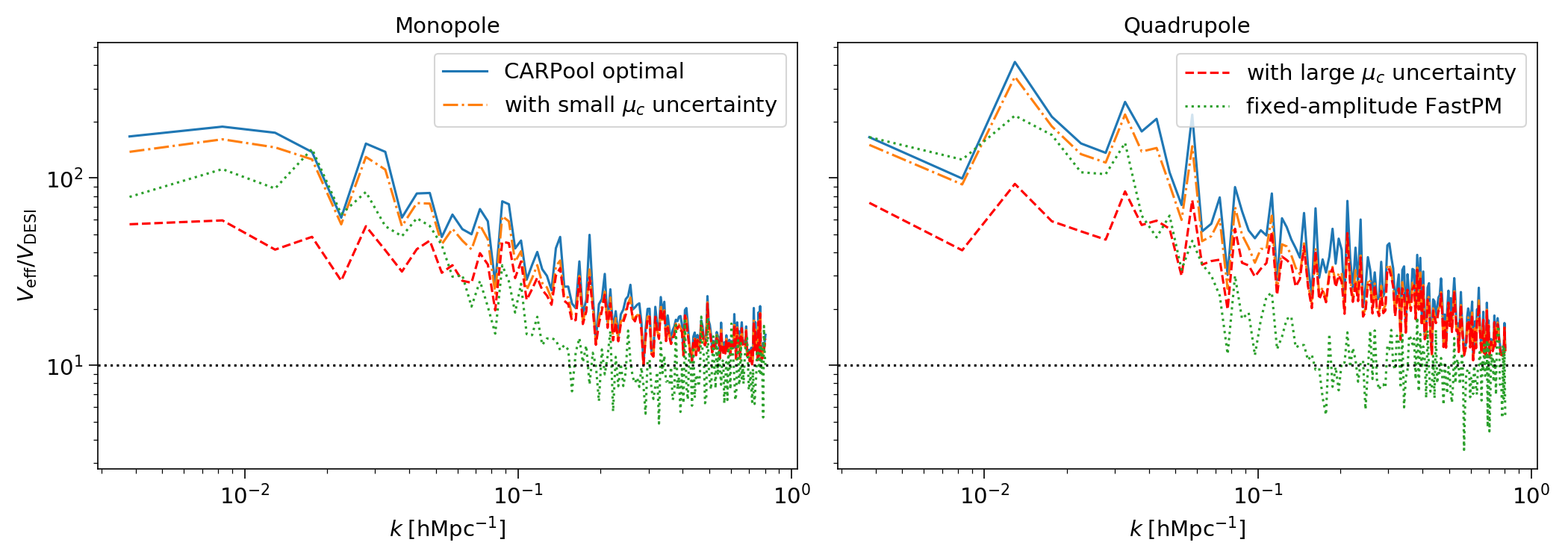}
    \caption{Same as Fig. \ref{fig:effective_volume} but for the halo catalogues with haloes massive than $10^{13}\Msunh$.}
    \label{fig:Veff_pk_mcut1e13}
\end{figure*}

We further study the CARPool performance on the halo catalogues with higher mass cut $10^{13}\Msunh$, which is about the mass boundary of host haloes of LRGs. Such catalogues are just the subsamples with high-mass haloes from the halo catalogues with mass cut $10^{11}\Msunh$. Fig. \ref{fig:Veff_pk_mcut1e13} shows the gain of effective volumes from CARPool for the monopole and quadrupole. Compared with the results from the halo catalogues with lower mass cut, the performance of CARPool is worse based on the increased effective volume. We think that it can be caused by shot noise, as the number of haloes decreases dramatically with higher mass cut. At lower redshifts ($z<1.1$), the number of LRG-host haloes is larger. For DESI, the number density of LRGs peaks around $z=0.6$--$0.8$. We check the CARPool performance at $z=0.8$ in Appendix \ref{appendix:z0.8}. Indeed, when the number density increases by $40\%$ from $z=1.1$ to $0.8$, the effective volume obtained from CARPool increases by $35\%$. Apart from that, the differences on the halo populations from the different halo finders, the non-linear structure growth and halo bias may affect the CARPool performance as well. We compare the Pearson correlation coefficients of the halo power spectrum multipoles from \textsc{AbacusSummit} and \textsc{FastPM} in Appendix \ref{sec:cov_pell}. The diagonal terms of the Pearson coefficients are closer to 1.0 for the lower halo mass cut, which indicates better performance from CARPool.

\subsection{Halo correlation function}
Since a correlation function is just the Fourier transform of power spectrum, we expect that the performance of CARPool on the halo correlation function is similar to the power spectrum. \cite{Chartier2021a} mainly studied the performance of CARPool on the clustering statistics in Fourier space, hence, it is worth investigating on the improvement in configuration space. For a cubic box, we can calculate the halo correlation function based on \cite{Peebles1974}, i.e.
\begin{align}
    \xi(s, \mu) = \frac{DD(s, \mu)}{RR(s, \mu)} - 1,
\end{align}
where $DD(s, \mu)$ and $RR(s, \mu)$ are respectively the normalized number of pairs of haloes and random points as a function of the separation distance $s=\sqrt{s_{\perp}^2 + s_{\|}^2}$ and the cosine angle between the separation vector $\bmath{s}$ and the line of sight, i.e. $\mu=s_{\|}/s$. We calculate the number of halo pairs using \textsc{Corrfunc}\footnote{https://github.com/manodeep/Corrfunc} \citep{Sinha2019, Sinha2020} and \textsc{FCFC}\footnote{https://github.com/cheng-zhao/FCFC} \citep{Zhao2021}, and have checked that the results from the two codes are consistent. For a cubic box, the number of random pairs can be predicted theoretically. In our calculation, we set $40$ linear radial bins in the range $5\leq s \leq 200 \Mpch$ and $60$ linear $\mu$ bins in the range $0 \leq \mu \leq 1.0$. Similar to the power spectrum multipoles, we calculate the correlation function multipoles from the anisotropic correlation function, i.e.
\begin{align}
    \xi_{\ell}(s) = (2\ell+1) \sum_{i=1}^{N_{\mu}}\xi(s, \mu_i) L_{\ell}(\mu_i)\Delta \mu,
\end{align}
where $N_{\mu}$ is the total number of $\mu$ bins.

\begin{figure*}
    \centering
    \includegraphics[width=1\linewidth]{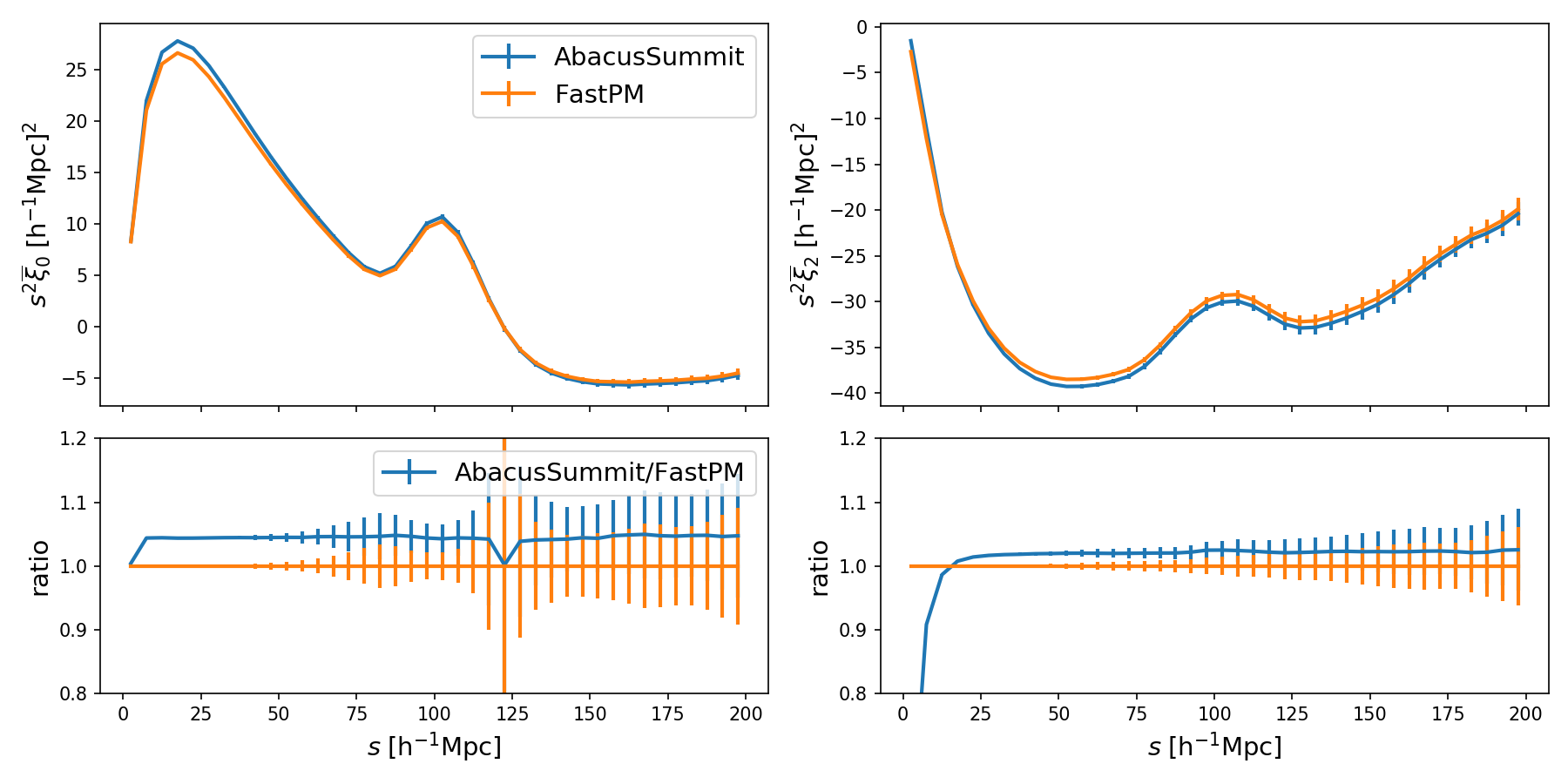}
    \caption{\textit{Upper panels}: the mean of the halo correlation function monopole (left-hand panel) and quadrupole (right-hand panel) from 25 paired \textsc{AbacusSummit} and \textsc{FastPM} catalogues with halo mass larger than $10^{11}\Msunh$ and at redshift $1.1$. The error bars have been scaled for the mean. The blue lines are the results of \textsc{AbacusSummit} and the orange lines are from \textsc{FastPM}. \textit{Lower panels}: the ratio of the mean between \textsc{AbacusSummit} and \textsc{FastPM}. The orange horizontal lines denote the results of \textsc{FastPM}. For the monopole ratio, the error bars have a sudden peak around $125\Mpch$, which is simply due to the zero crossing of the denominator.}
    \label{fig:xi_abacus_fastpm}
\end{figure*}

Fig. \ref{fig:xi_abacus_fastpm} shows the mean of the halo correlation function multipoles from 25 paired \textsc{AbacusSummit} and \textsc{FastPM} simulations, respectively. Same as Fig. \ref{fig:pk_mcut1.e11}, we study the halo catalogues at $z=1.1$ with mass cutoff $10^{11}\Msunh$. For the monopole, the difference from the two simulations is within $5\%$ over all the scales. For the quadrupole, the difference is less than that of the monopole, except for scales smaller than $10\Mpch$.

\subsubsection{Performance of CARPool method}
\begin{figure*}
    \centering
    \includegraphics[width=1\linewidth]{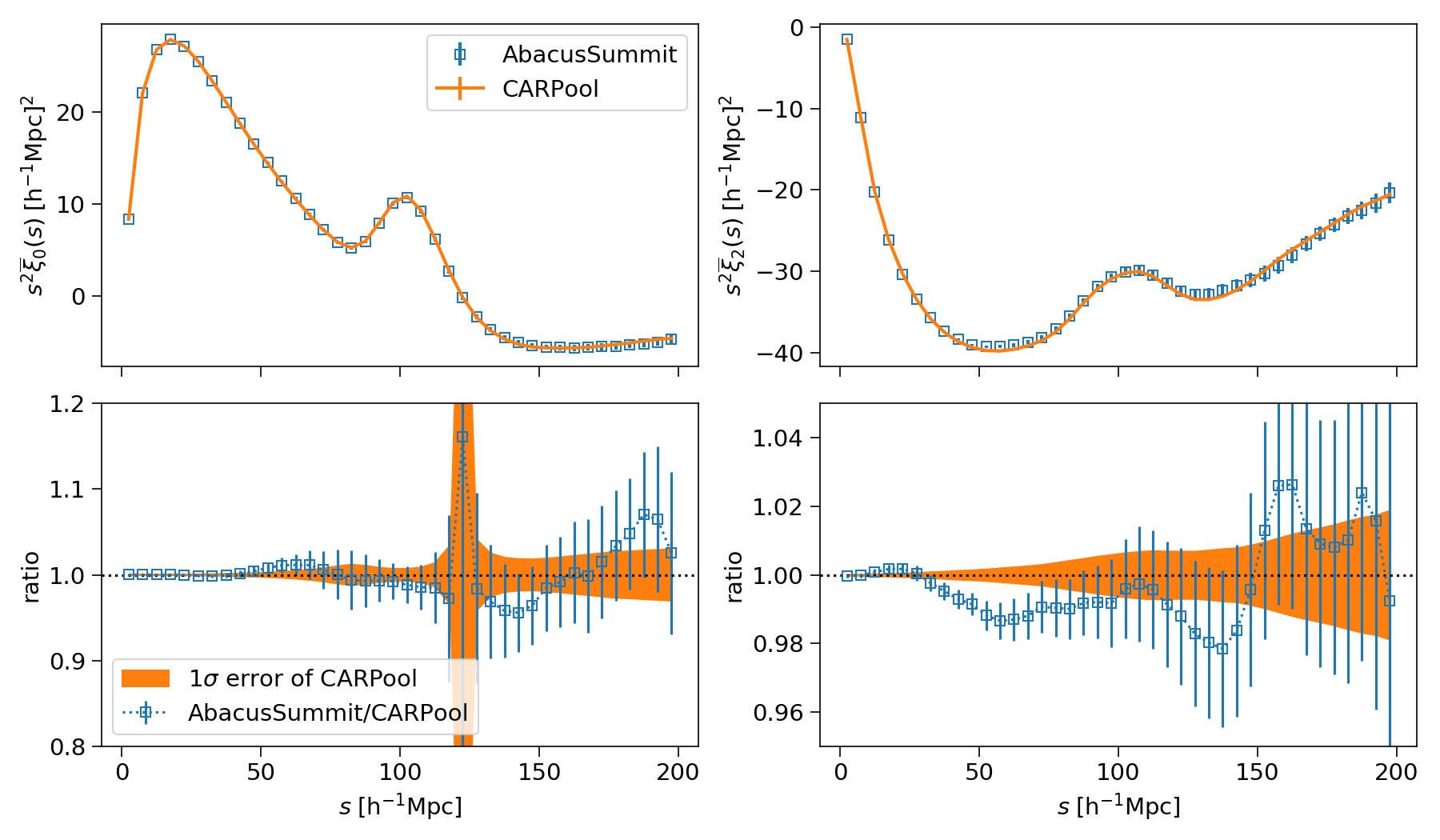}
    \caption{\textit{Upper panels}: the mean of the halo correlation function multipoles from 25 \textsc{AbacusSummit} simulations compared with the mean of $\xi_x$ from CARPool. The error bars have been scaled for the standard deviation of the mean multipoles. \textit{Lower panels}: the ratio of the mean before and after CARPool, shown as the blue points, along with $1\sigma$ error of the mean $\xi_x$, shown as the orange shaded regions. For the monopole ratio, the error bars have a sudden peak around $125\Mpch$, which is simply due to the zero crossing of the denominator. For the quadrupole, the discrepancy between the \textsc{AbacusSummit} and CARPool mean on scales around $50\Mpch$ is due to the statistical fluctuation of the mean from the paired \textsc{FastPM}, which we demonstrate in Fig. \ref{fig:xi2_fastpm_mcut1e11}.}
    \label{fig:xix_mcut1e11}
\end{figure*}

We show the halo correlation function multipoles from CARPool in Fig. \ref{fig:xix_mcut1e11}. Same as Fig. \ref{fig:px_mcut1e11}, we compare the mean of the monopoles (in the left-hand panels) and quadrupoles (in the right-hand panels) from \textsc{AbacusSummit} and CARPool. The upper panels show the overall shapes. The lower panels show the ratio of the mean multipoles of \textsc{AbacusSummit} over those of CARPool, given by the blue points. The shaded regions denote the standard deviations of the mean multipoles from CARPool. For the monopole, the relative difference of the mean is within $1\sigma$ error of \textsc{AbacusSummit}, while for the quadrupole, around scales $50\Mpch$, the difference is about $3\sigma$ error of \textsc{AbacusSummit}. However, it does not indicate that the CARPool result has bias; instead, it is just due to the sample variance from the paired \textsc{FastPM} catalogues, which we demonstrate in Appendix \ref{section:appendix_xi2}. 

Fig. \ref{fig:sigma_xix_mcut1e11} shows the reduction of the sample variance of the correlation function multipoles from CARPool. Similar to Fig. \ref{fig:sigma_pkx_mcut1e11}, we compare the cases whether the surrogate mean in CARPool is assumed as known or not. As shown in equation (\ref{eq:sigma_carpool_mean}), the uncertainty of the surrogate mean will propagate to that of CARPool result. For the red lines, we include the uncertainty of the mean correlation function multipoles that are calculated from the non-fixed-amplitude \textsc{FastPM} catalogues.   
Comparing Fig. \ref{fig:sigma_xix_mcut1e11} and \ref{fig:sigma_pkx_mcut1e11}, we see that the reduction of the sample variance from CARPool is consistent for the halo power spectrum and correlation function at large scales.

We also show the gain of the effective volume from CARPool based on the suppressed variance of the correlation function multipoles in Fig. \ref{fig:veff_xix}. The meaning of each line is the same as that with the same line type in Fig. \ref{fig:effective_volume}. The upper and lower panels show the results from the halo catalogues with mass cut $10^{11}$ and $10^{13}\Msunh$, respectively. We do not see the increase of $V_{\text{eff}}$ from small scales to large scales as what we observe in the case of the power spectrum multipoles. It is mainly due to the high cross-correlation between correlation function bins. 

\begin{figure*}
    \centering
    \includegraphics[width=1\linewidth]{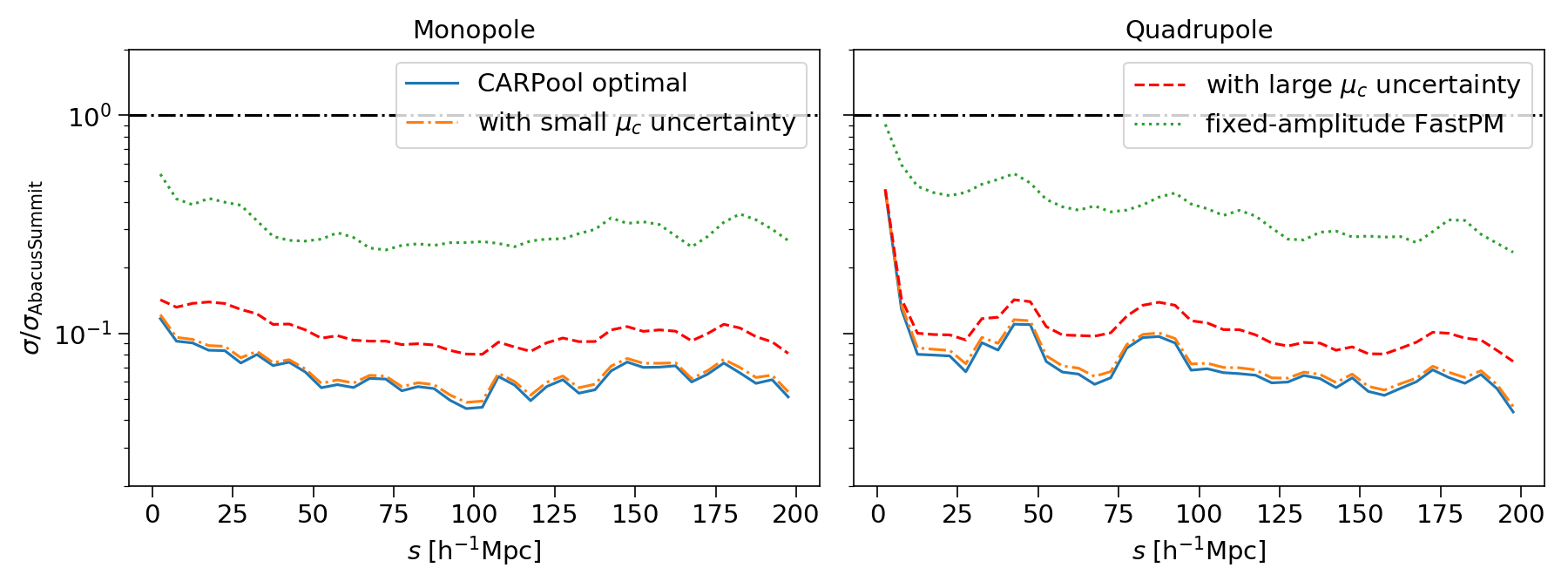}
    \caption{Similar to Fig. \ref{fig:sigma_pkx_mcut1e11} but for the correlation function multipoles. The blue lines stand for the optimal gains from CARPool assuming that the surrogate mean is known. The red dashed (orange dot--dashed) lines are the cases while considering the sample variance of the surrogate mean which are estimated from the non-fixed (fixed)-amplitude \textsc{FastPM}, respectively. The green dotted lines are obtained from the ratio of the standard deviation of the fixed-amplitude \textsc{FastPM} over that of \textsc{AbacusSummit}.}
    \label{fig:sigma_xix_mcut1e11}
\end{figure*}

\begin{figure*}
    \centering
    \includegraphics[width=1\linewidth]{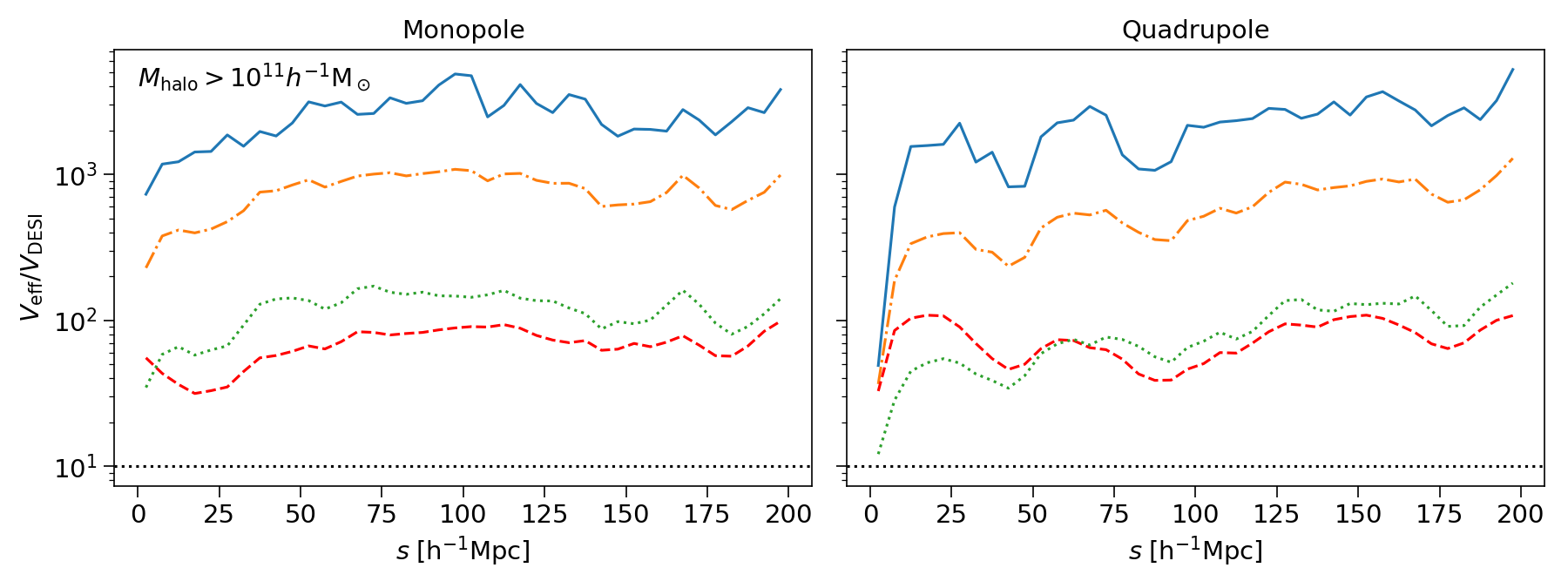}
    \includegraphics[width=1\linewidth]{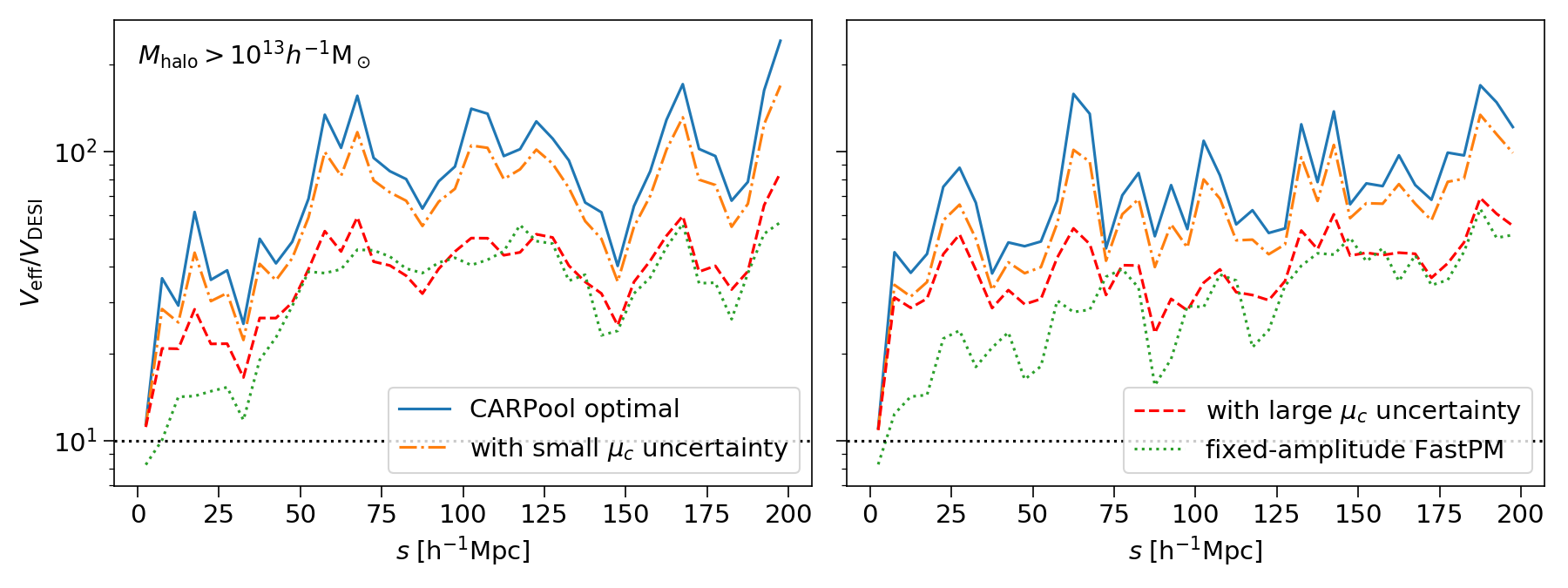}
    \caption{The effective volume derived from the halo correlation function multipoles using CARPool. We consider the effective volume from the combination of 25 \textsc{AbacusSummit} boxes. The upper (lower) panels show the results from the halo catalogues with halo mass larger than $10^{11}\Msunh$ ($10^{13}\Msunh$). The blue lines denote the optimal cases assuming that the surrogate mean is known, i.e. without considering the uncertainty of the surrogate mean. The red dashed (orange dot--dashed) lines take account of the statistical error of the surrogate mean estimated from the non-fixed (fixed)-amplitude \textsc{FastPM} catalogues. We also include the results if we only use 25 fixed-amplitude \textsc{FastPM} boxes, shown as the green dotted lines.}
    \label{fig:veff_xix}
\end{figure*}

\subsection{Halo Bispectrum}\label{sec:bispectrum}
\cite{Chartier2021a} have studied the performance of CARPool on matter bispectrum in real space and found similar significance as that of matter power spectrum. \cite{Chuang2019} studied the sample variance reduction from the fixed-amplitude method
on real-space halo bispectrum and found no improvement. Therefore,
it is interesting to check the CARPool performance on higher-order halo clustering statistics. We study the halo bispectrum defined as
\begin{align}
 \langle \delta(\bmath{k_1})\delta(\bmath{k_2})\delta(\bmath{k_3})\rangle \equiv \delta_D(\bmath{k_1} + \bmath{k_2} + \bmath{k_3}) B(k_1, k_2, k_3),
\end{align}
where $\delta(\bmath{k})$ is the halo number density contrast in Fourier space and $\delta_D$ is the Dirac delta function which ensures that the wavevectors $\bmath{k_1},\, \bmath{k_2}$ and $\bmath{k_3}$ form a closed triangle.

We use \textsc{Pylians3}\footnote{https://pylians3.readthedocs.io/en/master/index.html} \citep{Villaescusa-Navarro_2020} to calculate the bispectra for the halo catalogues with mass cut $10^{11}\Msunh$. The package implements a fast Fourier transform-based estimator \citep{Sefusatti_2005, Scoccimarro_2015, Sefusatti_2016} for the bispectrum calculation. The output bispectrum has been subtracted by the Poisson shot noise. Same as in \cite{Chuang2019}, we choose the triangle configuration of $k_1=0.1\hMpc$ and $k_2=0.2\hMpc$, which are the scales related to the baryon acoustic oscillations (BAO) and RSD analysis. We study the reduced bispectrum defined as
\begin{align}
    Q(k_1, k_2, k_3) = \frac{B(k_1, k_2, k_3)}{P(k_1)P(k_2) + P(k_1)P(k_3) + P(k_2)P(k_3)},
\end{align}
where $P(k)$ is the halo power spectrum.
Given the amplitude of $\bmath{k_1}$ and $\bmath{k_2}$, by varying $\bmath{k_3}$, the reduced bispectrum is a function of $\theta$, which is the angle between $\bmath{k_1}$ and $\bmath{k_2}$. We use $Q(\theta)$ to represent it.

\subsubsection{Performance of CARPool method}
\begin{figure}
    \centering
    \includegraphics[width=0.9\linewidth]{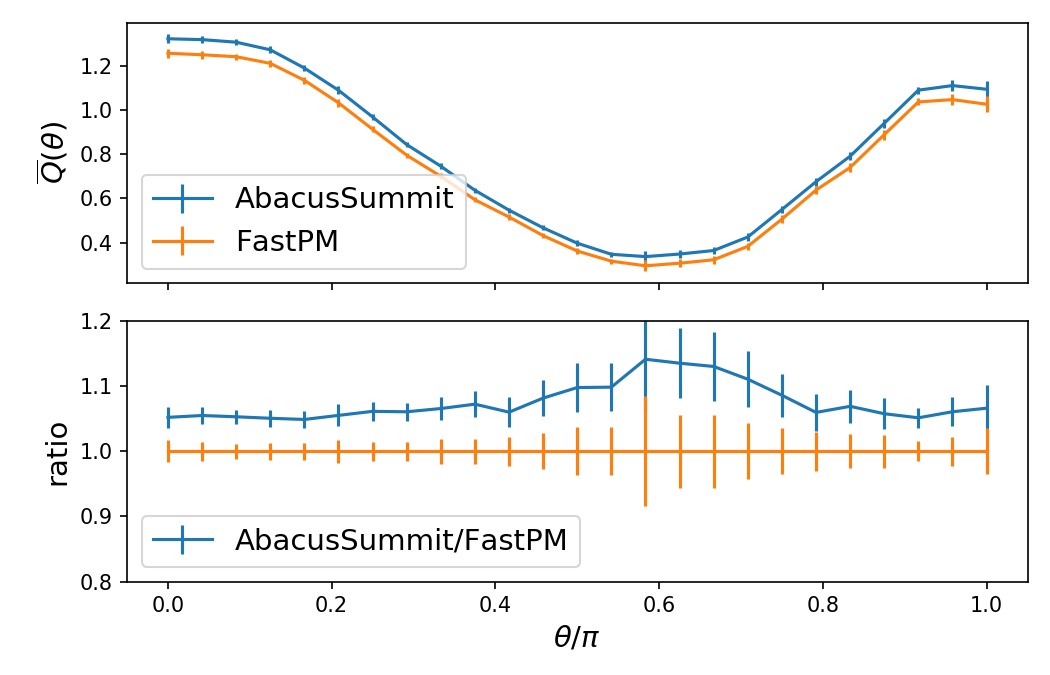}
    \includegraphics[width=0.9\linewidth]{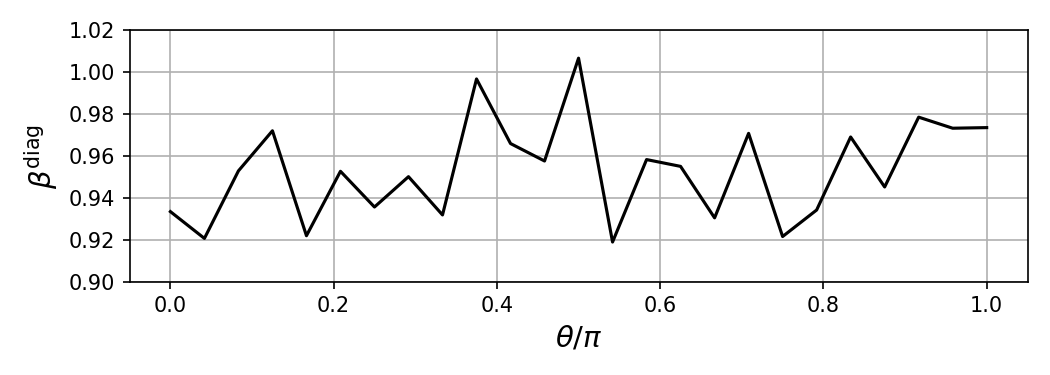}
    \includegraphics[width=0.9\linewidth]{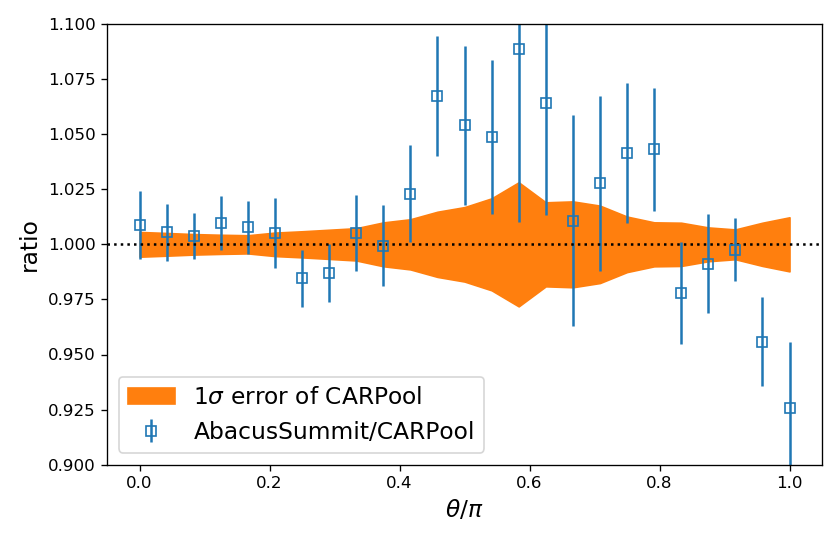}
    \caption{\textit{Top row}: the comparison of the real-space halo reduced bispectra from the \textsc{AbacusSummit} and \textsc{FastPM} halo catalogues with mass cut $10^{11}\Msunh$. The triangle configuration of the bispectrum is chosen with $k_1=0.1\hMpc$ and $k_2=0.2\hMpc$. The upper panel shows the mean of the monopoles averaged over 25 paired \textsc{AbacusSummit} and \textsc{FastPM} simulations, respectively. The lower panel shows the ratio between the two. \textit{Middle row}: the diagonal $\beta$ for the reduced bispectra calculated from equation (\ref{eq:beta_diag}). It is close to 1.0, indicating strong correlation between the bispectra from the \textsc{AbacusSummit} and \textsc{FastPM} catalogues. \textit{Bottom row}: the ratio between the mean of the reduced bispectra from \textsc{AbacusSummit} and CARPool, shown as the blue points. The shaded region shows the standard deviation of the CARPool result.}\label{fig:Bk}
\end{figure}

To compare the results in \cite{Chuang2019}, we show our results in real space, but we have checked that the results in redshift space are similar too. Same as the analysis in halo power spectrum, we first compare the mean of the reduced bispectra from 25 paired \textsc{AbacusSummit} and \textsc{FastPM} halo catalogues, shown in the first row of Fig. \ref{fig:Bk}. We see that the bispectra from the two sets of halo catalogues agree relatively well with each other. The difference is within $10\%$. We also check the $\beta^{\text{diag}}$ which is based on equation (\ref{eq:beta_diag}) and find it close to 1.0, shown as the middle row. We calculate the mean of the bispectra from the non-fixed-amplitude \textsc{FastPM} catalogues. We implement all the necessary elements into the equation of CARPool and obtain the final result. We compare the mean of CARPool and that of \textsc{AbacusSummit} and show the ratio between the two in the bottom row of the figure. The error bars denote the standard deviation of the mean of \textsc{AbacusSummit}, and the shaded region is the standard deviation of the CARPool mean with the consideration of the error of the surrogate mean from the non-fixed-amplitude \textsc{FastPM} catalogues. The agreement between the mean $Q(\theta)$ from CARPool and \textsc{AbacusSummit} is within $3\sigma$ error of \textsc{AbacusSummit}.

\begin{figure}
    \centering
    \includegraphics[width=0.9\linewidth]{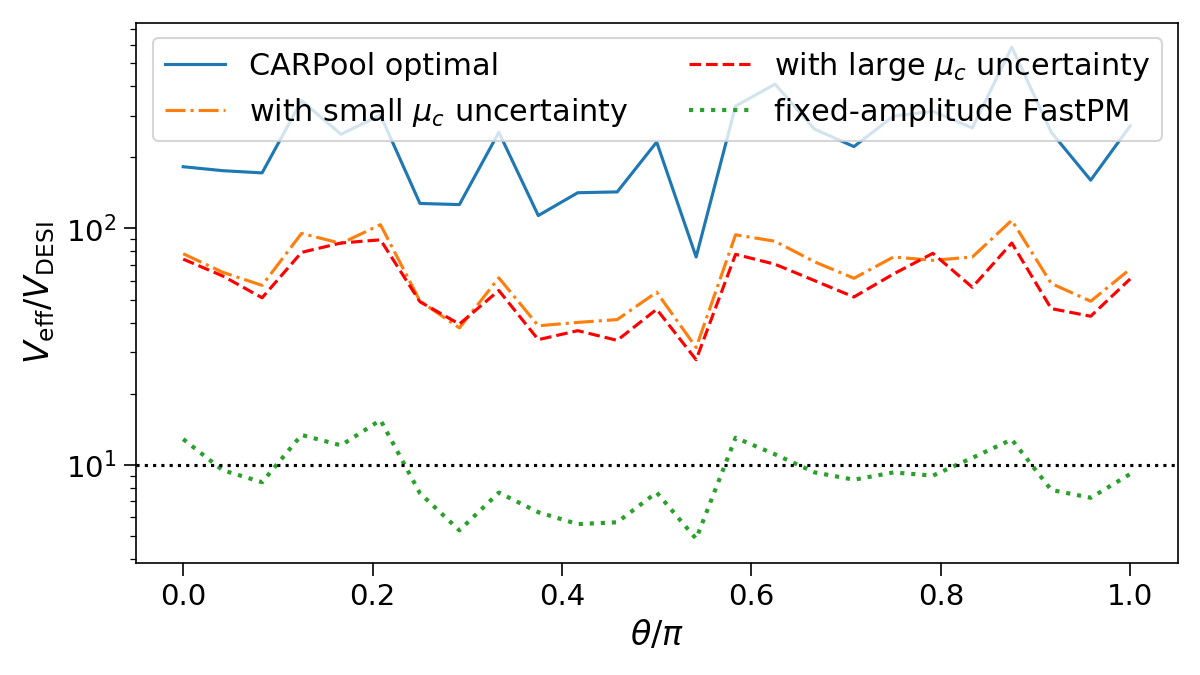}
    \caption{Similar to Fig. \ref{fig:effective_volume}, we show the increased effective volume for the reduced bispectrum from CARPool. The blue line is the optimal case. The red dashed line and the orange dot--dashed line represent the cases taking account of the uncertainty of the mean bispectra estimated from the non-fixed-amplitude and the fixed-amplitude \textsc{FastPM} catalogues, respectively. The green dotted line represents the result if we only use the fixed-amplitude \textsc{FastPM} catalogues.}
    \label{fig:Veff_Bk}
\end{figure}

Comparing the standard deviation of the reduced bispectrum from CARPool and \textsc{AbacusSummit}, we can obtain the increased effective volume due to CARPool. Similar to Fig. \ref{fig:effective_volume}, we show the effective volume from 25 \textsc{AbacusSummit} boxes before and after CARPool in Fig. \ref{fig:Veff_Bk}. Again, the volume of 25 \textsc{AbacusSummit} base boxes corresponds to 10 times the DESI volume, shown as the horizontal dotted line. CARPool can also significantly increase the effective volume for the halo bispectrum. The blue line shows the optimal case without considering the error of the surrogate mean. The red dashed line denotes a realistic case considering the error of the surrogate mean from the non-fixed-amplitude \textsc{FastPM} catalogues. It is about five times larger than the default volume.

We also check the improvement from the fixed-amplitude method, based on the standard deviation of the reduced bispectrum from the fixed-amplitude \textsc{FastPM} catalogues. The result is shown as the green dotted line, matching with the black dotted line, indicating no improvement, which is consistent with the finding of \cite{Chuang2019}. Therefore, even if we use the surrogate mean from the fixed-amplitude catalogues instead of the non-fixed-amplitude ones in CARPool, there will be little improvement, shown as the orange dot--dashed line. \cite{Angulo2016} found that the paired-and-fixed method can suppress the sample variance of the matter bispectrum with the triangle configuration $k_1=0.02\hMpc$ and $k_2=0.04\hMpc$ which is at a larger scale than the one we study here.
But the variance suppression mainly comes from pairing instead of fixing.
\cite{Klypin2020} further checked that as the scale becomes smaller, around the scales of BAO, the reduction from the paired-and-fixed method quickly vanishes. We believe that it should be true for the halo bispectrum too. At least, we have checked that the fixed-amplitude method does not reduce the sample variance of the halo bispectrum at the configuration of $k_1=0.02\hMpc$ and $k_2=0.04\hMpc$.

The failure of variance suppression from the fixed-amplitude method can be understood from the theoretical investigations on the bispectrum in literature.
\cite{Matsubara2007} and \cite{Qin2021} found that the phases of the non-Gaussian density field are vital to the bispectrum. The late-time evolution induced phase autocorrelation and phase-modulus cross-correlation contribute almost equally to the bispectrum, whereas the modulus autocorrelation contributes little. Therefore, fixing the amplitude (modulus) of the initial density field should not affect the sample variance of the bispectrum. Adding pairing with the inverse phases may help reducing the sample variance from the phase-modulus cross-correlation at very large scales, but the reduction quickly vanishes as the scale becomes smaller. 
On the contrary, CARPool shows significant advantage in reducing the sample variance of bispectrum.
For more detailed study on the bispectrum from the fixed-amplitude method (with pairing) and CARPool, we leave it for future work.

\section{Application on Simulations with Different Cosmologies }\label{sec:diff_cosmo}
We extend the CARPool method to different cosmologies, even if there is only one \textsc{AbacusSummit} simulation for a given cosmology. 
For secondary cosmologies, \textsc{AbacusSummit} has much fewer number of simulations compared with that of the base cosmology. For the base box size, there are only six simulations for some of the secondary cosmologies (i.e. c$00[2-4]$), and only one simulation for each of the other secondary cosmologies. If we use the same CARPool method as that in the base cosmology, there are two bottlenecks. One is that we do not have enough \textsc{AbacusSummit} simulations to pair with \textsc{FastPM} in order to calculate the cross-correlation between the two. The other is that we need to generate a large number of \textsc{FastPM} simulations to estimate the mean of clustering statistics, which takes a larger amount of efforts as well. To resolve these, we modify the CARPool method by using the products in the base cosmology.

Since for the base cosmology we have the initial white noise of the 25 \textsc{AbacusSummit} boxes, we can use them to generate the same number of \textsc{FastPM} simulations for a given secondary cosmology. We can pair these \textsc{FastPM} simulations with the \textsc{AbacusSummit} ones in the base cosmology. Since they share the same initial white noise, there is non-zero cross-correlation between the halo statistics from the two sets of simulations even though with different cosmologies. We calculate the cross-correlation between the \textsc{AbacusSummit} from the base cosmology and the \textsc{FastPM} from a second cosmology. We assume that it is close to the cross-correlation between the \textsc{AbacusSummit} from the secondary cosmology and the \textsc{FastPM} from the base cosmology, i.e.
\begin{align}
  \text{Cov}(A_{\text{c002}}, F_{\text{c000}}) \approx \text{Cov}(A_{\text{c000}}, F_{\text{c002}}). \label{eq:cov_c002_c000}
\end{align}
where $A$ and $F$ are short for \textsc{AbacusSummit} and \textsc{FastPM}, respectively. The subscript c002 represents a secondary cosmology as an example. We roughly examine the viability of the above relation in Appendix \ref{section:appendix_cross_correlation}, but a better validation can be conducted if we have $25$ sets of \textsc{AbacusSummit} simulations with the secondary cosmology c002. In the end, we construct the CARPool result for a secondary cosmology via
\begin{align}
    X_{\text{c002}} = A_{\text{c002}} - \beta (F_{\text{c000}} - \overline{F}^{\prime}_{\text{c000}}), \label{eq:px_diff_cosmo}
\end{align}
where 
\begin{align}
    \beta = \text{Diag}[\text{Cov}(A_{\text{c000}}, F_{\text{c002}})]/\text{Var}(F_{\text{c000}}).\label{eq:beta_diff_cosmo}
\end{align}
Note that in the denominator of $\beta$ we still use the variance of the \textsc{FastPM} statistics from c000 instead of c002, so that we only modify the cross-correlation term (numerator) compared with the exact definition of $\beta$. Another approach we are developing to estimate the cross-correlation is using the jackknife method (see Zhang et al., in preparation).

\begin{figure*}
    \centering
    \includegraphics[width=1.0\linewidth]{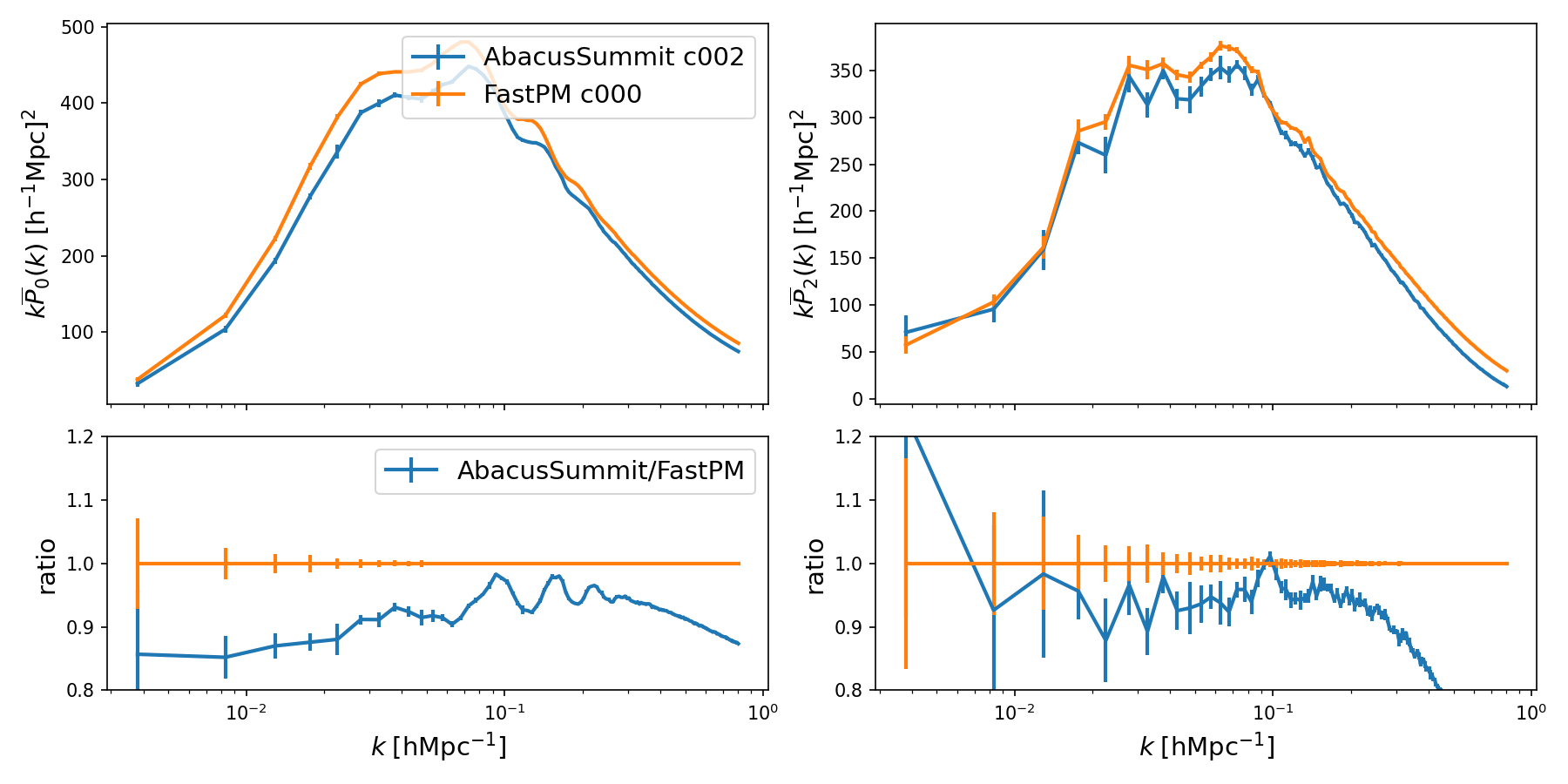}
    \caption{The comparison of the mean power spectrum multipoles from the \textsc{AbacusSummit} with the secondary cosmology c002 and from the \textsc{FastPM} with the base cosmology c000. For the \textsc{AbacusSummit} with c002, there are only 6 boxes, while for the \textsc{FastPM} with c000, there are 25 boxes. The standard deviations of the \textsc{FastPM} multipoles are smaller than those of the \textsc{AbacusSummit}.}
    \label{fig:pk_c002}
\end{figure*}

We apply our method for the secondary cosmology c002. We first compare the mean of the halo power spectrum multipoles from the \textsc{AbacusSummit} halo catalogues with c002 and from the \textsc{FastPM} halo catalogues with c000 in Fig. \ref{fig:pk_c002}. The halo catalogues are at $z=1.1$ with halo mass larger than $10^{11}\Msunh$. The left-hand panels are for the monopoles and the right-hand panels are for the quadrupoles. For the \textsc{AbacusSummit} with c002, there are only 6 boxes, while for the \textsc{FastPM} we include all the 25 boxes with the same white noise from the \textsc{AbacusSummit} in c000, so that the statistical noise of the \textsc{AbacusSummit} mean is larger than that of the \textsc{FastPM}. In the left-hand panels, we show the ratio of the mean multipoles from the two simulations, denoted as the blue lines. For the monopole, there is BAO residual (with wiggles) caused by the shifting of $k$ coordinates due to the difference of the dark energy parameters $w_0$ and $w_a$ between the two cosmologies.  

\begin{figure}
    \centering
    \includegraphics[width=1.0\linewidth]{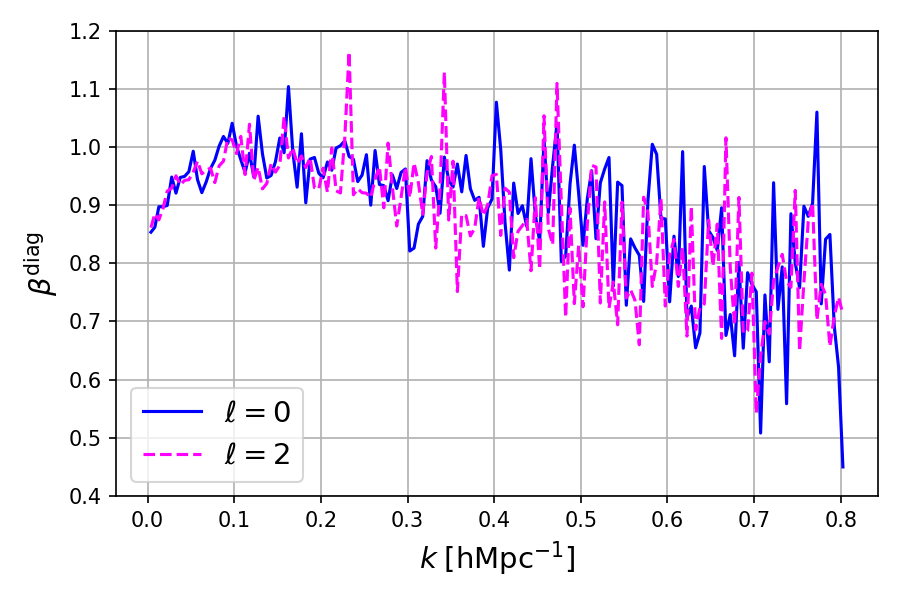}
    \caption{Similar as Fig. \ref{fig:beta_pk_mcut1e11}, but the $\beta^{\text{diag}}$ is calculated from the cross-correlation between the halo power spectrum multipoles from the paired \textsc{AbacusSummit} in c000 and the \textsc{FastPM} in c002.}
    \label{fig:beta_c000_c002}
\end{figure}
We show the diagonal terms of $\beta$ for the halo power spectrum monopole and quadrupole calculated from equation (\ref{eq:beta_diff_cosmo}) in Fig. \ref{fig:beta_c000_c002}. The general shapes are similar to the ones in Fig. \ref{fig:beta_pk_mcut1e11}.
\begin{figure*}
    \centering
    \includegraphics[width=1.0\linewidth]{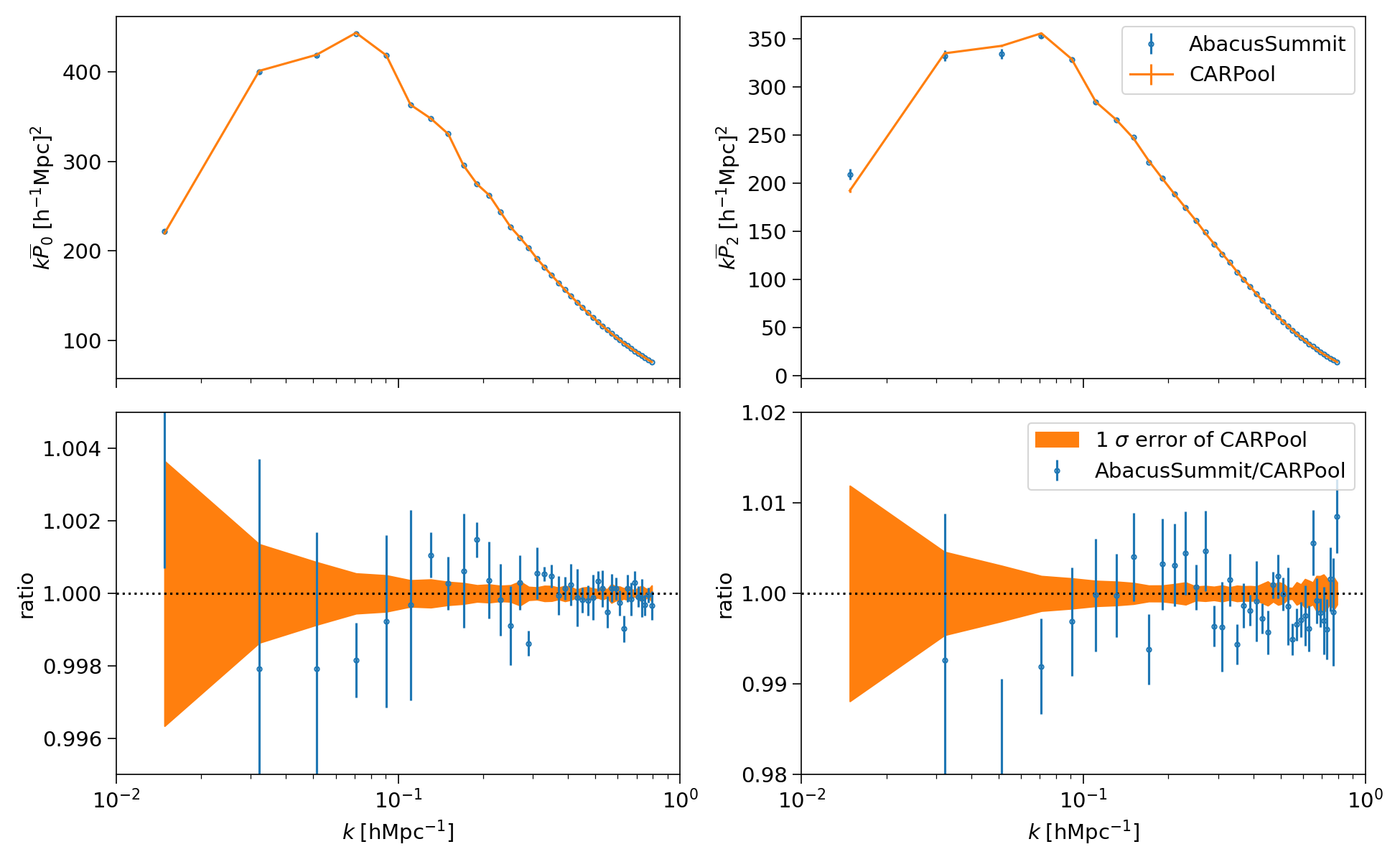}
    \caption{Same as Fig. \ref{fig:px_mcut1e11} but for the halo power spectrum multipoles from the secondary cosmology c002. We have also rebinned $k$ to have better clarity for the data points.}
    \label{fig:px_c002}
\end{figure*}
We show the results of the multipoles from CARPool in Fig. \ref{fig:px_c002}. Same as Fig. \ref{fig:px_mcut1e11}, the blue points are the mean from six \textsc{AbacusSummit} halo catalogues. The orange lines and shaded regions represent the CARPool results. From the ratio between the mean, on the one hand, we do not see any bias of the CARPool result compared with the \textsc{AbacusSummit}; on the other hand the variance reduction from CARPool is also significant. Therefore, the method of CARPool also applies for the clustering from a different cosmology. 

\begin{figure*}
    \centering
    \includegraphics[width=1.0\linewidth]{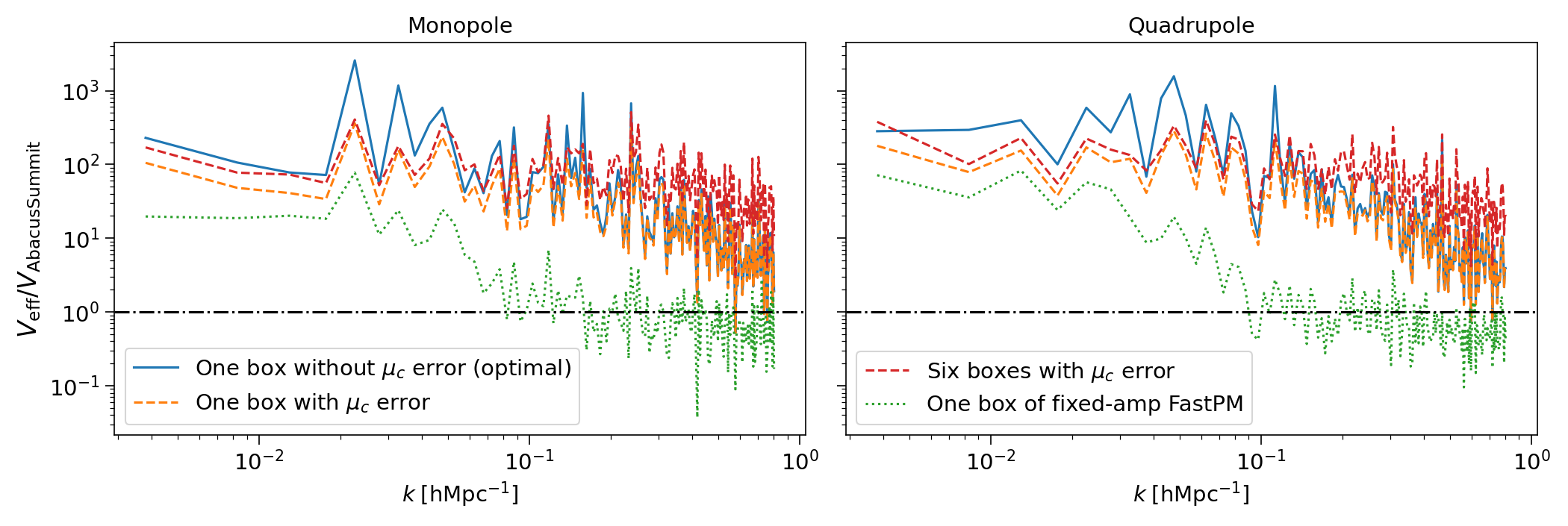}
    \caption{The gain on the effective volume for the halo power spectrum multipoles with the secondary cosmology c002 from CARPool. The left-hand panel is for the monopole and the right-hand panel is for the quadrupole. We show the increase of effective volume in terms of the volume of one \textsc{AbacusSummit} box. In each panel, the blue line denotes the effective volume of one \textsc{AbacusSummit} box with CARPool assuming that the surrogate mean is known. The orange dashed line is the same as the blue line but includes the statistical error of the surrogate mean estimated from the non-fixed-amplitude \textsc{FastPM} catalogues. Similar to the orange dashed line, the red dashed line is the effective volume if we combine six \textsc{AbacusSummit} boxes with CARPool. The green dotted line denotes the effective volume from one fixed-amplitude \textsc{FastPM} box.}
    \label{fig:veff_c002}
\end{figure*}

To see the gain of the effective volume from CARPool compared with the default volume of one \textsc{AbacusSummit} box, we study the situations from one box and from the combination of six boxes, respectively, as shown in Fig. \ref{fig:veff_c002}. In each panel, the blue line represents the effective volume of one \textsc{AbacusSummit} box from CARPool with the assumption that the surrogate mean is known. For the realistic case where we estimate the surrogate mean from the non-fixed-amplitude \textsc{FastPM} catalogues with c000, the effective volume decreases by about half on large scales shown as the orange dashed line. On scales up to $k=0.3\hMpc$, the gain of the effective volume is above $20$ times. For the effective volume from the fixed-amplitude method shown as the green dotted line, it has similar but lower gain compared with the CARPool results, and it quickly decreases to zero at smaller scales $k>0.1\hMpc$. We also check the effective volume from the combination of six boxes with CARPool shown as the red dashed line, which has noticeable increase over all the scales compared to the one box case.

We perform the same analysis for the secondary cosmology c004 and find similar results. We show the results in Appendix \ref{appendix:carpool_c004}. 

\section{Conclusions and Discussions}\label{sec:conclusion}
We have prepared a set of \textsc{FastPM} simulations
including 25 boxes with the \textsc{AbacusSummit} ICs,
>200 boxes with independent ICs,
237 boxes with the fixed-amplitude ICs,
and 2 sets of 25 boxes with different cosmologies.

We use the \textsc{FastPM} simulations to improve the precision of \textsc{AbacusSummit} simulations by adopting the CARPool method. We study the clustering of two halo catalogues at $z=1.1$, (i) halo mass $> 10^{11}~h^{-1}$M$_{\sun}$ (i.e. hosts of DESI ELGs) and (ii) halo mass $> 10^{13}~h^{-1}$M$_{\sun}$ (i.e. hosts of DESI LRGs). We present the sample variance reduction and the increased effective volume from CARPool. For the ELG-host-halo catalogues, the effective volume is larger than 100 times DESI volume, i.e. $2000 \Gpchcube$ at $k<0.3\hMpc$ for the power spectrum measurements. We confirm that the fixed-amplitude variance suppressing method can be very effective at smaller $k$ (e.g. $k < 0.1\hMpc$) but the performance drops quickly while $k$ increases (e.g. no improvement at $k > 0.2\hMpc$). CARPool performs better than the fixed-amplitude method at smaller scales. We also check the effective volume for the correlation function. It is larger than 30 times DESI volume at $r>10\Mpch$ from CARPool. Similarly, for the LRG-host-halo catalogues, the effective volume is larger than 20 times DESI volume (i.e. $400\Gpchcube$) at $k < 0.2\hMpc$ and larger than 20 times DESI volume at $r > 10\Mpch$.

Furthermore, CARPool can effectively suppress the sample variance of the halo bispectrum at the region of BAO scales, whereas the fixed-amplitude method cannot, which is consistent with the findings in \cite{Chuang2019, Klypin2020}. Using CARPool, we can obtain larger than 50 times DESI volume for the ELG-host-halo bispectrum with the triangle configuration $k_1=0.1\hMpc$ and $k_2=0.2\hMpc$. 
The trivial gain from the fixed-amplitude method is probably due to the fact that the bispectrum is not mainly contributed from the modulus correlation of the density field, but the phase correlation and phase-modulus cross-correlation \citep{Matsubara2007, Qin2021}.

We further generalize the method to increase the effective volume of simulations with different cosmologies. These simulations have only one or a few simulations for each cosmology while there are 25 simulations for the primary cosmology. Even with one simulation, we find that the effective volume can be increased by more than 20 times. This finding can be useful when we want to understand how our data analysis pipeline responds to different cosmologies. We can then avoid the systematic bias introduced by a fixed fiducial cosmology model. On the other hand, a tricky step for the generalized method is to estimate the cross-covariance between an \textsc{AbacusSummit} simulation and the paired \textsc{FastPM} simulation with different cosmologies. We have used an approximation in this paper but we are developing another approach based on an internal sample evaluation (Zhang et al., in preparation).

The performance of this technique depends on the cross-correlation between the paired $N$-body and approximate simulations. There are many methods producing approximate simulations, e.g. see \cite{Chuang2015} and \cite{lippich2019comparing} for various methods, but we do not expect that the other methods can perform better than quasi-$N$-body codes, e.g. \textsc{FastPM}, \textsc{COLA} \citep{Tassev2013}, etc. On the other hand, the computing time for generating quasi-$N$-body simulations is still not negligible; thus, it could be interesting to try other more efficient methods generating approximate simulations.

We have shown that massive approximate gravity calculations combined with a limited number of accurate $N$-body simulations can be exploited with the CARPool and variance suppression techniques to obtain accurate error estimates on the two-point and three-point statistics of halo clustering. For future work, we can extend CARPool to \textsc{AbacusSummit} galaxy mocks constructed for DESI observables. CARPool can help to improve the constraints on galaxy clustering models which are related with cosmological parameters and galaxy--halo connection. Since we do not expect a good one-to-one match between the low-mass haloes in \textsc{AbacusSummit} and \textsc{FastPM} runs, the galaxy assignment on \textsc{FastPM} side may not provide much gains on the cross-correlation.
We are expecting that the stochasticity in the galaxy--halo connection might reduce the performance of CARPool at small scales. 
On the other hand, we expect that CARPool is likely to perform better on the constraints of galaxy--halo connection than the fixed-amplitude method (with pairing), since the latter one fails to reduce the sample variance of halo power spectrum at $k>0.2\hMpc$ \citep[also see][]{Avila2022} and halo bispectrum.
All in all, it is worth investigating on CARPool with galaxy clustering in the next step. This work sets the path for a robust cosmological analysis of galaxy surveys.

\section*{Acknowledgements}
We thank the DESI internal reviewers Francisco-Shu Kitaura and Shadab Alam for their constructed comments. We appreciate that the referee give many insightful suggestions and comments to improve the manuscript.
ZD thanks Haojie Xu, Hai Yu, Pengjie Zhang and Zhao Chen for helpful discussions.
ZD and YY were supported by the National Key Basic Research and Development Program of China (No. 2018YFA0404504) and the National Science Foundation of China (grant nos. 11621303, 11890691, 11773048).
LHG is supported by the Center for Computational Astrophysics at the Flatiron Institute, which is supported by the Simons Foundation.  \ABACUS development has been supported by NSF AST-1313285 and DOE-SC0013718, as well as by Harvard University start-up funds.\\

 This research is supported by the Director, Office of Science, Office of High Energy Physics of the U.S. Department of Energy under Contract No. DE–AC02–05CH11231, and by the National Energy Research Scientific Computing Center, a DOE Office of Science User Facility under the same contract; additional support for DESI is provided by the U.S. National Science Foundation, Division of Astronomical Sciences under Contract No. AST-0950945 to the NSF’s National Optical-Infrared Astronomy Research Laboratory; the Science and Technologies Facilities Council of the United Kingdom; the Gordon and Betty Moore Foundation; the Heising-Simons Foundation; the French Alternative Energies and Atomic Energy Commission (CEA); the National Council of Science and Technology of Mexico; the Ministry of Economy of Spain, and by the DESI Member Institutions.
We are honoured to be permitted to conduct scientific research on Iolkam Du’ag (Kitt Peak), a mountain with particular significance to the Tohono O’odham Nation. 

We produced \textsc{FastPM} simulations at the National Energy Research Scientific Computing Center (NERSC). We thank Stephen Bailey for providing the computational resources from the DESI quota. We thank NERSC staff for their highly responsive and expert assistance, both scientific and administrative, during the course of this project. Part of the clustering measurements were performed at the Gravity Supercomputer in the Department of Astronomy, Shanghai Jiao Tong University.

\section*{Data Availability}

The \textsc{AbacusSummit} simulations used in this study are publicly available (https://abacusnbody.org/). We plan to make the \textsc{FastPM} simulations available along with one of the DESI data releases. The halo clustering statistics used in this study, including correlation functions, power spectra and bispectra from the \textsc{AbacusSummit} and \textsc{FastPM} simulations, is available at https://doi.org/10.5281/zenodo.5993283.

\bibliographystyle{mnras}
\bibliography{references}

\begin{thebibliography}{}
\makeatletter
\relax
\def\mn@urlcharsother{\let\do\@makeother \do\$\do\&\do\#\do\^\do\_\do\%\do\~}
\def\mn@doi{\begingroup\mn@urlcharsother \@ifnextchar [ {\mn@doi@}
  {\mn@doi@[]}}
\def\mn@doi@[#1]#2{\def\@tempa{#1}\ifx\@tempa\@empty \href
  {http://dx.doi.org/#2} {doi:#2}\else \href {http://dx.doi.org/#2} {#1}\fi
  \endgroup}
\def\mn@eprint#1#2{\mn@eprint@#1:#2::\@nil}
\def\mn@eprint@arXiv#1{\href {http://arxiv.org/abs/#1} {{\tt arXiv:#1}}}
\def\mn@eprint@dblp#1{\href {http://dblp.uni-trier.de/rec/bibtex/#1.xml}
  {dblp:#1}}
\def\mn@eprint@#1:#2:#3:#4\@nil{\def\@tempa {#1}\def\@tempb {#2}\def\@tempc
  {#3}\ifx \@tempc \@empty \let \@tempc \@tempb \let \@tempb \@tempa \fi \ifx
  \@tempb \@empty \def\@tempb {arXiv}\fi \@ifundefined
  {mn@eprint@\@tempb}{\@tempb:\@tempc}{\expandafter \expandafter \csname
  mn@eprint@\@tempb\endcsname \expandafter{\@tempc}}}

\bibitem[\protect\citeauthoryear{{Alam} et~al.,}{{Alam}
  et~al.}{2021}]{Alam_2020}
{Alam} S.,  et~al., 2021, \mn@doi [\jcap] {10.1088/1475-7516/2021/11/050},
  \href {https://ui.adsabs.harvard.edu/abs/2021JCAP...11..050A} {2021, 050}

\bibitem[\protect\citeauthoryear{{Allison}, {Caucal}, {Calabrese}, {Dunkley}
  \& {Louis}}{{Allison} et~al.}{2015}]{Allison_2015}
{Allison} R.,  {Caucal} P.,  {Calabrese} E.,  {Dunkley} J.,   {Louis} T.,
  2015, \mn@doi [\prd] {10.1103/PhysRevD.92.123535}, \href
  {https://ui.adsabs.harvard.edu/abs/2015PhRvD..92l3535A} {92, 123535}

\bibitem[\protect\citeauthoryear{{Angulo} \& {Pontzen}}{{Angulo} \&
  {Pontzen}}{2016}]{Angulo2016}
{Angulo} R.~E.,  {Pontzen} A.,  2016, \mn@doi [\mnras] {10.1093/mnrasl/slw098},
  \href {https://ui.adsabs.harvard.edu/abs/2016MNRAS.462L...1A} {462, L1}

\bibitem[\protect\citeauthoryear{{Avila} \& {Gutierrez Adame}}{{Avila} \&
  {Gutierrez Adame}}{2022}]{Avila2022}
{Avila} S.,  {Gutierrez Adame} A.,  2022, arXiv e-prints, \href
  {https://ui.adsabs.harvard.edu/abs/2022arXiv220411103A} {p. arXiv:2204.11103}

\bibitem[\protect\citeauthoryear{{Avila} et~al.,}{{Avila}
  et~al.}{2020}]{Avila2020}
{Avila} S.,  et~al., 2020, \mn@doi [\mnras] {10.1093/mnras/staa2951}, \href
  {https://ui.adsabs.harvard.edu/abs/2020MNRAS.499.5486A} {499, 5486}

\bibitem[\protect\citeauthoryear{Avramidis \& Wilson}{Avramidis \&
  Wilson}{1993}]{Avramidis1993}
Avramidis A.~N.,  Wilson J.~R.,  1993, \mn@doi [Operations Research Letters]
  {https://doi.org/10.1016/0167-6377(93)90069-S}, 14, 187

\bibitem[\protect\citeauthoryear{{Bayer}, {Banerjee}  \& {Seljak}}{{Bayer}
  et~al.}{2021a}]{bayer2021fake}
{Bayer} A.~E.,  {Banerjee} A.,   {Seljak} U.,  2021a, arXiv e-prints, \href
  {https://ui.adsabs.harvard.edu/abs/2021arXiv210804215B} {p. arXiv:2108.04215}

\bibitem[\protect\citeauthoryear{{Bayer} et~al.,}{{Bayer}
  et~al.}{2021b}]{bayer2021detecting}
{Bayer} A.~E.,  et~al., 2021b, \mn@doi [\apj] {10.3847/1538-4357/ac0e91}, \href
  {https://ui.adsabs.harvard.edu/abs/2021ApJ...919...24B} {919, 24}

\bibitem[\protect\citeauthoryear{{Bayer}, {Banerjee}  \& {Feng}}{{Bayer}
  et~al.}{2021c}]{Bayer2020}
{Bayer} A.~E.,  {Banerjee} A.,   {Feng} Y.,  2021c, \mn@doi [\jcap]
  {10.1088/1475-7516/2021/01/016}, \href
  {https://ui.adsabs.harvard.edu/abs/2021JCAP...01..016B} {2021, 016}

\bibitem[\protect\citeauthoryear{{Bose}, {Eisenstein}, {Hadzhiyska}, {Garrison}
   \& {Yuan}}{{Bose} et~al.}{2022}]{Bose2021}
{Bose} S.,  {Eisenstein} D.~J.,  {Hadzhiyska} B.,  {Garrison} L.~H.,   {Yuan}
  S.,  2022, \mn@doi [\mnras] {10.1093/mnras/stac555}, \href
  {https://ui.adsabs.harvard.edu/abs/2022MNRAS.512..837B} {512, 837}

\bibitem[\protect\citeauthoryear{{Chartier}, {Wandelt}, {Akrami}  \&
  {Villaescusa-Navarro}}{{Chartier} et~al.}{2021}]{Chartier2021a}
{Chartier} N.,  {Wandelt} B.,  {Akrami} Y.,   {Villaescusa-Navarro} F.,  2021,
  \mn@doi [\mnras] {10.1093/mnras/stab430}, \href
  {https://ui.adsabs.harvard.edu/abs/2021MNRAS.503.1897C} {503, 1897}

\bibitem[\protect\citeauthoryear{Chuang et~al.,}{Chuang
  et~al.}{2015}]{Chuang2015}
Chuang C.-H.,  et~al., 2015, \mn@doi [Monthly Notices of the Royal Astronomical
  Society] {10.1093/mnras/stv1289}, 452, 686–700

\bibitem[\protect\citeauthoryear{{Chuang} et~al.,}{{Chuang}
  et~al.}{2019}]{Chuang2019}
{Chuang} C.-H.,  et~al., 2019, \mn@doi [\mnras] {10.1093/mnras/stz1233}, \href
  {https://ui.adsabs.harvard.edu/abs/2019MNRAS.487...48C} {487, 48}

\bibitem[\protect\citeauthoryear{{Copeland}, {Sami}  \& {Tsujikawa}}{{Copeland}
  et~al.}{2006}]{Copeland_2006}
{Copeland} E.~J.,  {Sami} M.,   {Tsujikawa} S.,  2006, \mn@doi [International
  Journal of Modern Physics D] {10.1142/S021827180600942X}, \href
  {https://ui.adsabs.harvard.edu/abs/2006IJMPD..15.1753C} {15, 1753}

\bibitem[\protect\citeauthoryear{{DESI Collaboration} et~al.,}{{DESI
  Collaboration} et~al.}{2016}]{DESIscience2016}
{DESI Collaboration} et~al., 2016, arXiv e-prints, \href
  {https://ui.adsabs.harvard.edu/abs/2016arXiv161100036D} {p. arXiv:1611.00036}

\bibitem[\protect\citeauthoryear{{Feng}, {Chu}, {Seljak}  \& {McDonald}}{{Feng}
  et~al.}{2016}]{Feng2016}
{Feng} Y.,  {Chu} M.-Y.,  {Seljak} U.,   {McDonald} P.,  2016, \mn@doi [\mnras]
  {10.1093/mnras/stw2123}, \href
  {https://ui.adsabs.harvard.edu/abs/2016MNRAS.463.2273F} {463, 2273}

\bibitem[\protect\citeauthoryear{{Font-Ribera}, {McDonald}, {Mostek}, {Reid},
  {Seo}  \& {Slosar}}{{Font-Ribera} et~al.}{2014}]{DESInu}
{Font-Ribera} A.,  {McDonald} P.,  {Mostek} N.,  {Reid} B.~A.,  {Seo} H.-J.,
  {Slosar} A.,  2014, \mn@doi [Journal of Cosmology and Astroparticle Physics]
  {10.1088/1475-7516/2014/05/023}, \href
  {http://adsabs.harvard.edu/abs/2014JCAP...05..023F} {5, 023}

\bibitem[\protect\citeauthoryear{{Garrison}, {Eisenstein}, {Ferrer}, {Metchnik}
   \& {Pinto}}{{Garrison} et~al.}{2016}]{Garrison2016}
{Garrison} L.~H.,  {Eisenstein} D.~J.,  {Ferrer} D.,  {Metchnik} M.~V.,
  {Pinto} P.~A.,  2016, \mn@doi [\mnras] {10.1093/mnras/stw1594}, \href
  {https://ui.adsabs.harvard.edu/abs/2016MNRAS.461.4125G} {461, 4125}

\bibitem[\protect\citeauthoryear{{Garrison}, {Eisenstein}, {Ferrer}, {Tinker},
  {Pinto}  \& {Weinberg}}{{Garrison} et~al.}{2018}]{Garrison2018}
{Garrison} L.~H.,  {Eisenstein} D.~J.,  {Ferrer} D.,  {Tinker} J.~L.,  {Pinto}
  P.~A.,   {Weinberg} D.~H.,  2018, \mn@doi [\apjs] {10.3847/1538-4365/aabfd3},
  \href {https://ui.adsabs.harvard.edu/abs/2018ApJS..236...43G} {236, 43}

\bibitem[\protect\citeauthoryear{{Garrison}, {Eisenstein}  \&
  {Pinto}}{{Garrison} et~al.}{2019}]{Garrison2019}
{Garrison} L.~H.,  {Eisenstein} D.~J.,   {Pinto} P.~A.,  2019, \mn@doi [\mnras]
  {10.1093/mnras/stz634}, \href
  {https://ui.adsabs.harvard.edu/abs/2019MNRAS.485.3370G} {485, 3370}

\bibitem[\protect\citeauthoryear{{Garrison}, {Eisenstein}, {Ferrer},
  {Maksimova}  \& {Pinto}}{{Garrison} et~al.}{2021}]{Garrison2021}
{Garrison} L.~H.,  {Eisenstein} D.~J.,  {Ferrer} D.,  {Maksimova} N.~A.,
  {Pinto} P.~A.,  2021, \mn@doi [\mnras] {10.1093/mnras/stab2482}, \href
  {https://ui.adsabs.harvard.edu/abs/2021MNRAS.508..575G} {508, 575}

\bibitem[\protect\citeauthoryear{{Gonzalez-Perez} et~al.,}{{Gonzalez-Perez}
  et~al.}{2018}]{Gonzalez2018}
{Gonzalez-Perez} V.,  et~al., 2018, \mn@doi [\mnras] {10.1093/mnras/stx2807},
  \href {https://ui.adsabs.harvard.edu/abs/2018MNRAS.474.4024G} {474, 4024}

\bibitem[\protect\citeauthoryear{{Grove} et~al.,}{{Grove}
  et~al.}{2021}]{Grove2021}
{Grove} C.,  et~al., 2021, arXiv e-prints, \href
  {https://ui.adsabs.harvard.edu/abs/2021arXiv211209138G} {p. arXiv:2112.09138}

\bibitem[\protect\citeauthoryear{{Hadzhiyska}, {Eisenstein}, {Bose}, {Garrison}
   \& {Maksimova}}{{Hadzhiyska} et~al.}{2022}]{Hadzhiyska2021}
{Hadzhiyska} B.,  {Eisenstein} D.,  {Bose} S.,  {Garrison} L.~H.,   {Maksimova}
  N.,  2022, \mn@doi [\mnras] {10.1093/mnras/stab2980}, \href
  {https://ui.adsabs.harvard.edu/abs/2022MNRAS.509..501H} {509, 501}

\bibitem[\protect\citeauthoryear{{Hahn} \& {Villaescusa-Navarro}}{{Hahn} \&
  {Villaescusa-Navarro}}{2021}]{hahn2020constraining}
{Hahn} C.,  {Villaescusa-Navarro} F.,  2021, \mn@doi [\jcap]
  {10.1088/1475-7516/2021/04/029}, \href
  {https://ui.adsabs.harvard.edu/abs/2021JCAP...04..029H} {2021, 029}

\bibitem[\protect\citeauthoryear{{Hand}, {Feng}, {Beutler}, {Li}, {Modi},
  {Seljak}  \& {Slepian}}{{Hand} et~al.}{2018}]{Hand2018}
{Hand} N.,  {Feng} Y.,  {Beutler} F.,  {Li} Y.,  {Modi} C.,  {Seljak} U.,
  {Slepian} Z.,  2018, \mn@doi [\aj] {10.3847/1538-3881/aadae0}, \href
  {https://ui.adsabs.harvard.edu/abs/2018AJ....156..160H} {156, 160}

\bibitem[\protect\citeauthoryear{{Hern{\'a}ndez-Aguayo}, {Prada}, {Baugh}  \&
  {Klypin}}{{Hern{\'a}ndez-Aguayo} et~al.}{2021}]{Hernandez2021}
{Hern{\'a}ndez-Aguayo} C.,  {Prada} F.,  {Baugh} C.~M.,   {Klypin} A.,  2021,
  \mn@doi [\mnras] {10.1093/mnras/stab434}, \href
  {https://ui.adsabs.harvard.edu/abs/2021MNRAS.503.2318H} {503, 2318}

\bibitem[\protect\citeauthoryear{{Huterer} et~al.,}{{Huterer}
  et~al.}{2015}]{Huterer_2015}
{Huterer} D.,  et~al., 2015, \mn@doi [Astroparticle Physics]
  {10.1016/j.astropartphys.2014.07.004}, \href
  {https://ui.adsabs.harvard.edu/abs/2015APh....63...23H} {63, 23}

\bibitem[\protect\citeauthoryear{{Klypin}, {Prada}  \& {Byun}}{{Klypin}
  et~al.}{2020}]{Klypin2020}
{Klypin} A.,  {Prada} F.,   {Byun} J.,  2020, \mn@doi [\mnras]
  {10.1093/mnras/staa734}, \href
  {https://ui.adsabs.harvard.edu/abs/2020MNRAS.496.3862K} {496, 3862}

\bibitem[\protect\citeauthoryear{{Kreisch}, {Pisani}, {Villaescusa-Navarro},
  {Spergel}, {Wandelt}, {Hamaus}  \& {Bayer}}{{Kreisch}
  et~al.}{2021}]{Kreisch_2021}
{Kreisch} C.~D.,  {Pisani} A.,  {Villaescusa-Navarro} F.,  {Spergel} D.~N.,
  {Wandelt} B.~D.,  {Hamaus} N.,   {Bayer} A.~E.,  2021, arXiv e-prints, \href
  {https://ui.adsabs.harvard.edu/abs/2021arXiv210702304K} {p. arXiv:2107.02304}

\bibitem[\protect\citeauthoryear{Lesgourgues \& Pastor}{Lesgourgues \&
  Pastor}{2006}]{LESGOURGUES_2006}
Lesgourgues J.,  Pastor S.,  2006, \mn@doi [Physics Reports]
  {10.1016/j.physrep.2006.04.001}, 429, 307–379

\bibitem[\protect\citeauthoryear{Lippich et~al.,}{Lippich
  et~al.}{2019}]{lippich2019comparing}
Lippich M.,  et~al., 2019, Monthly Notices of the Royal Astronomical Society,
  482, 1786

\bibitem[\protect\citeauthoryear{{Maion}, {Angulo}  \& {Zennaro}}{{Maion}
  et~al.}{2022}]{Maion2022}
{Maion} F.,  {Angulo} R.~E.,   {Zennaro} M.,  2022, arXiv e-prints, \href
  {https://ui.adsabs.harvard.edu/abs/2022arXiv220403868M} {p. arXiv:2204.03868}

\bibitem[\protect\citeauthoryear{{Maksimova}, {Garrison}, {Eisenstein},
  {Hadzhiyska}, {Bose}  \& {Satterthwaite}}{{Maksimova}
  et~al.}{2021}]{Maksimova2021}
{Maksimova} N.~A.,  {Garrison} L.~H.,  {Eisenstein} D.~J.,  {Hadzhiyska} B.,
  {Bose} S.,   {Satterthwaite} T.~P.,  2021, \mn@doi [\mnras]
  {10.1093/mnras/stab2484}, \href
  {https://ui.adsabs.harvard.edu/abs/2021MNRAS.508.4017M} {508, 4017}

\bibitem[\protect\citeauthoryear{{Massara}, {Villaescusa-Navarro}, {Ho},
  {Dalal}  \& {Spergel}}{{Massara} et~al.}{2021}]{Massara_2020}
{Massara} E.,  {Villaescusa-Navarro} F.,  {Ho} S.,  {Dalal} N.,   {Spergel}
  D.~N.,  2021, \mn@doi [\prl] {10.1103/PhysRevLett.126.011301}, \href
  {https://ui.adsabs.harvard.edu/abs/2021PhRvL.126a1301M} {126, 011301}

\bibitem[\protect\citeauthoryear{{Matsubara}}{{Matsubara}}{2007}]{Matsubara2007}
{Matsubara} T.,  2007, \mn@doi [\apjs] {10.1086/513466}, \href
  {https://ui.adsabs.harvard.edu/abs/2007ApJS..170....1M} {170, 1}

\bibitem[\protect\citeauthoryear{{Metchnik}}{{Metchnik}}{2009}]{Metchnik2009}
{Metchnik} M. V.~L.,  2009, PhD thesis, The University of Arizona

\bibitem[\protect\citeauthoryear{{Peebles} \& {Hauser}}{{Peebles} \&
  {Hauser}}{1974}]{Peebles1974}
{Peebles} P.~J.~E.,  {Hauser} M.~G.,  1974, \mn@doi [\apjs] {10.1086/190308},
  \href {https://ui.adsabs.harvard.edu/abs/1974ApJS...28...19P} {28, 19}

\bibitem[\protect\citeauthoryear{{Planck Collaboration} et~al.,}{{Planck
  Collaboration} et~al.}{2020}]{Planck2018}
{Planck Collaboration} et~al., 2020, \mn@doi [\aap]
  {10.1051/0004-6361/201833910}, \href
  {https://ui.adsabs.harvard.edu/abs/2020A&A...641A...6P} {641, A6}

\bibitem[\protect\citeauthoryear{{Pontzen}, {Slosar}, {Roth}  \&
  {Peiris}}{{Pontzen} et~al.}{2016}]{Pontzen2016}
{Pontzen} A.,  {Slosar} A.,  {Roth} N.,   {Peiris} H.~V.,  2016, \mn@doi [\prd]
  {10.1103/PhysRevD.93.103519}, \href
  {https://ui.adsabs.harvard.edu/abs/2016PhRvD..93j3519P} {93, 103519}

\bibitem[\protect\citeauthoryear{{Porta Nova} \& {Wilson}}{{Porta Nova} \&
  {Wilson}}{1993}]{deOPortaNova1993}
{Porta Nova} A.~M.,  {Wilson} J.~R.,  1993, \mn@doi [European Journal of
  Operational Research] {https://doi.org/10.1016/0377-2217(93)90262-L}, 71, 80

\bibitem[\protect\citeauthoryear{Qin, Pan, Yu  \& Zhang}{Qin
  et~al.}{2022}]{Qin2021}
Qin J.,  Pan J.,  Yu Y.,   Zhang P.,  2022, \mn@doi [\mnras]
  {10.1093/mnras/stac1454}, 514, 1548

\bibitem[\protect\citeauthoryear{{Quinn}, {Katz}, {Stadel}  \& {Lake}}{{Quinn}
  et~al.}{1997}]{Quinn1997}
{Quinn} T.,  {Katz} N.,  {Stadel} J.,   {Lake} G.,  1997, arXiv e-prints, \href
  {https://ui.adsabs.harvard.edu/abs/1997astro.ph.10043Q} {pp
  astro--ph/9710043}

\bibitem[\protect\citeauthoryear{Rubinstein \& Marcus}{Rubinstein \&
  Marcus}{1985}]{Rubinstein1985}
Rubinstein R.~Y.,  Marcus R.,  1985, \mn@doi [Operations Research]
  {10.1287/opre.33.3.661}, 33, 661

\bibitem[\protect\citeauthoryear{{Scoccimarro}}{{Scoccimarro}}{2015}]{Scoccimarro_2015}
{Scoccimarro} R.,  2015, \mn@doi [\prd] {10.1103/PhysRevD.92.083532}, \href
  {https://ui.adsabs.harvard.edu/abs/2015PhRvD..92h3532S} {92, 083532}

\bibitem[\protect\citeauthoryear{{Sefusatti} \& {Scoccimarro}}{{Sefusatti} \&
  {Scoccimarro}}{2005}]{Sefusatti_2005}
{Sefusatti} E.,  {Scoccimarro} R.,  2005, \mn@doi [\prd]
  {10.1103/PhysRevD.71.063001}, \href
  {https://ui.adsabs.harvard.edu/abs/2005PhRvD..71f3001S} {71, 063001}

\bibitem[\protect\citeauthoryear{{Sefusatti}, {Crocce}, {Scoccimarro}  \&
  {Couchman}}{{Sefusatti} et~al.}{2016}]{Sefusatti_2016}
{Sefusatti} E.,  {Crocce} M.,  {Scoccimarro} R.,   {Couchman} H.~M.~P.,  2016,
  \mn@doi [\mnras] {10.1093/mnras/stw1229}, \href
  {https://ui.adsabs.harvard.edu/abs/2016MNRAS.460.3624S} {460, 3624}

\bibitem[\protect\citeauthoryear{{Sinha} \& {Garrison}}{{Sinha} \&
  {Garrison}}{2019}]{Sinha2019}
{Sinha} M.,  {Garrison} L.~H.,  2019, in Software Challenges to Exascale
  Computing. Second Workshop. pp 3--20 (\mn@eprint {arXiv} {1911.08275}),
  \mn@doi{10.1007/978-981-13-7729-7\_1}

\bibitem[\protect\citeauthoryear{{Sinha} \& {Garrison}}{{Sinha} \&
  {Garrison}}{2020}]{Sinha2020}
{Sinha} M.,  {Garrison} L.~H.,  2020, \mn@doi [\mnras] {10.1093/mnras/stz3157},
  \href {https://ui.adsabs.harvard.edu/abs/2020MNRAS.491.3022S} {491, 3022}

\bibitem[\protect\citeauthoryear{{Tassev}, {Zaldarriaga}  \&
  {Eisenstein}}{{Tassev} et~al.}{2013}]{Tassev2013}
{Tassev} S.,  {Zaldarriaga} M.,   {Eisenstein} D.~J.,  2013, \mn@doi [\jcap]
  {10.1088/1475-7516/2013/06/036}, \href
  {https://ui.adsabs.harvard.edu/abs/2013JCAP...06..036T} {2013, 036}

\bibitem[\protect\citeauthoryear{{Villaescusa-Navarro}
  et~al.,}{{Villaescusa-Navarro} et~al.}{2018}]{Villaescusa2018}
{Villaescusa-Navarro} F.,  et~al., 2018, \mn@doi [\apj]
  {10.3847/1538-4357/aae52b}, \href
  {https://ui.adsabs.harvard.edu/abs/2018ApJ...867..137V} {867, 137}

\bibitem[\protect\citeauthoryear{{Villaescusa-Navarro}
  et~al.,}{{Villaescusa-Navarro} et~al.}{2020}]{Villaescusa-Navarro_2020}
{Villaescusa-Navarro} F.,  et~al., 2020, \mn@doi [\apjs]
  {10.3847/1538-4365/ab9d82}, \href
  {https://ui.adsabs.harvard.edu/abs/2020ApJS..250....2V} {250, 2}

\bibitem[\protect\citeauthoryear{{Zhao} et~al.,}{{Zhao}
  et~al.}{2021}]{Zhao2021}
{Zhao} C.,  et~al., 2021, \mn@doi [\mnras] {10.1093/mnras/stab510}, \href
  {https://ui.adsabs.harvard.edu/abs/2021MNRAS.503.1149Z} {503, 1149}

\bibitem[\protect\citeauthoryear{{Zhou} et~al.,}{{Zhou}
  et~al.}{2021}]{Zhou2021}
{Zhou} R.,  et~al., 2021, \mn@doi [\mnras] {10.1093/mnras/staa3764}, \href
  {https://ui.adsabs.harvard.edu/abs/2021MNRAS.501.3309Z} {501, 3309}

\makeatother
\end{thebibliography}


\appendix

\section{Configuration parameters of the \textsc{FastPM} simulation}\label{sec:appendix_fastpm_parameters}
Before the massive production of the \textsc{FastPM} simulations, we run some tests on the main configuration parameters, i.e. the force resolution parameter B, the time-steps T, as well as the redshift of the IC. We would like to find a set of configuration parameters that can make the \textsc{FastPM} simulation as close to the \textsc{AbacusSummit} simulation as possible, based on their matter power spectra. Meanwhile, such configuration should not be expensive to run, since even for \textsc{FastPM}, running a $2\Gpch$ box with mass resolution $5\times 10^9\Msunh$ still takes considerable computational resources. 

\begin{figure}
    \centering
    \includegraphics[width=1.0\linewidth]{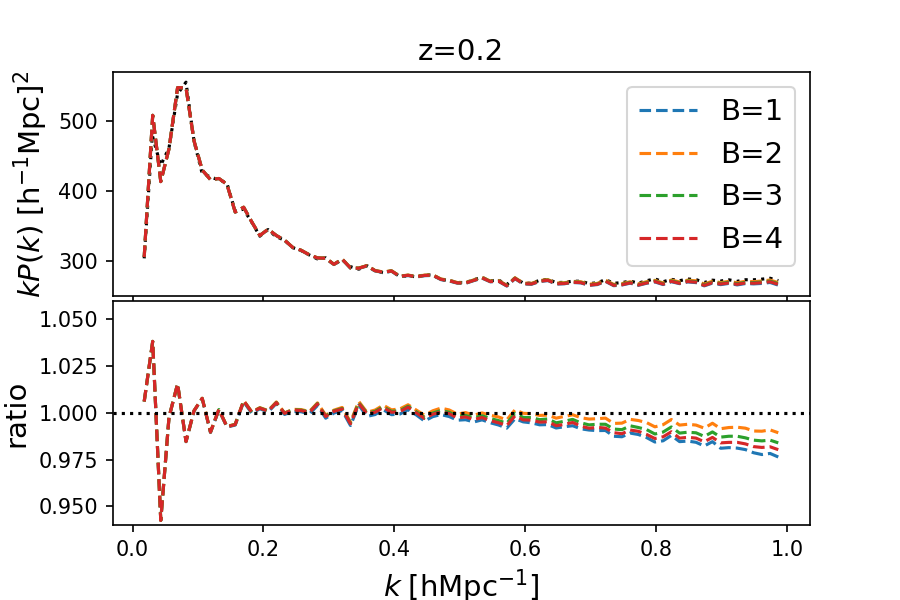}
    \caption{The comparison between the matter power spectra at $z=0.2$ from the \textsc{AbacusSummit} (black dotted line) and the \textsc{FastPM} (coloured lines) with different force resolution parameters $B$s but the same $T=40$ and $a_0=0.05$. The upper panel shows the overall shapes up to $k=1.0\hMpc$. The lower panel shows the ratio between the \textsc{FastPM} and the \textsc{AbacusSummit} which we take as the standard. Different $B$s cause the deviations of \textsc{FastPM} on small scales differently. In this case, $B=2$ performs the best. }\label{fig:diffB_fastpm}
\end{figure}

\begin{figure}
    \centering
    \includegraphics[width=1.0\linewidth]{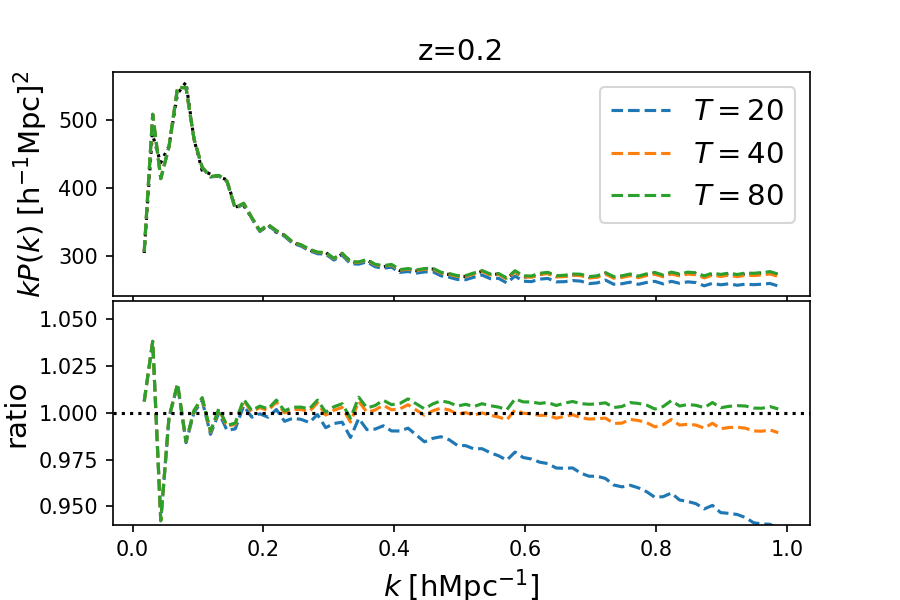}
    \caption{Similar to Fig. \ref{fig:diffB_fastpm} but for the test of the number of time-steps $T$. We fix $B=2$ and $a_0=0.05$. Considering the accuracy and computation time, $T=40$ performs better than $T=20$ and $80$. }\label{fig:diffT_fastpm}
\end{figure}

\begin{figure}
    \centering
    \includegraphics[width=1.0\linewidth]{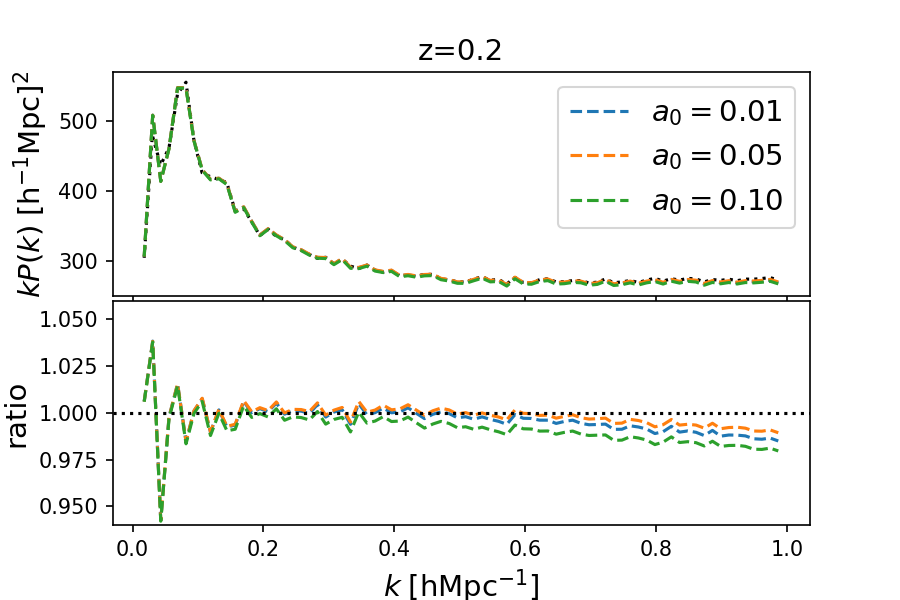}
    \caption{Similar to Fig. \ref{fig:diffB_fastpm} but for the test of the initial scale $a_0$. We fix $B=2$ and $T=40$. The initial scale $a_0=0.05$, i.e. $z_0=19$, performs better than the other two.}\label{fig:diffa0_fastpm}
\end{figure}

In fact, \textsc{AbacusSummit} has generated many simulations with box size $500\Mpch$, which can be used to check the \textsc{FastPM} performance. Using the IC from one such \textsc{AbacusSummit} box, we generate different \textsc{FastPM} simulations by varying the configuration parameters. To reduce the sample variance, we fix the amplitude of the IC in the \textsc{FastPM} simulations. We set the same particle mass resolution as that of the \textsc{FastPM} base box ($2\Gpch$). Thanks to the small box size, it is relatively cheap and convenient to generate many realizations for the test purpose. We believe that a set of reasonable configuration parameters we find for a small box still holds for a larger box with the same particle mass resolution. 
To determine the \textsc{FastPM} performance from a given set of configuration parameters, we compare the real-space matter power spectrum with that of \textsc{AbacusSummit} at redshift $z=0.2$. We show the comparison in Fig. \ref{fig:diffB_fastpm}--\ref{fig:diffa0_fastpm}. In each figure, we vary one parameter and fix the other two. 

In Fig. \ref{fig:diffB_fastpm}, we show the impact from B while fixing $T=40$ and the initial redshift $z_0=19$, i.e. the initial scale $a_0=0.05$. Note that in our case T is always counted as the number of time-steps from a given initial redshift to the finial redshift 0.1 with the linear step size in scale $a$. Different coloured lines represent the results from different $B$s. The black dotted line denotes the result from \textsc{AbacusSummit}, which we take as the standard. In the upper panel, we show the matter power spectra at $z=0.2$ from \textsc{AbacusSummit} as well as from \textsc{FastPM} with different $B$s. In the lower panel, we show the ratio between the power spectra from \textsc{FastPM} and \textsc{AbacusSummit}. Since the embedded routines for calculating matter power spectra in \textsc{FastPM} and \textsc{AbacusSummit} are not exactly the same, the number of modes from some $k$ bins varies a bit due to different counting schemes for the $k$ modes on the bin boundaries, hence, it can cause some fluctuation at very large scales where the number of modes is small. We can ignore such fluctuation from numerical issues. Based on the ratio, we see the deviation of the \textsc{FastPM} power spectra from the \textsc{AbacusSummit} on small scales. The deviation from $B=1$ is the largest, about $2.5\%$, as it has the lowest force resolution. Interestingly, $B=2$ gives the lowest deviation (about $1\%$ at $k=1.0\hMpc$) than the cases with $B=3$ or $4$. The reason is beyond our knowledge, which we leave it for future study. 

In Fig. \ref{fig:diffT_fastpm}, we vary $T=20$, $40$, $80$, but fix $B=2$ and $a_0=0.05$. The performance from $T=40$ is comparable with that from $T=80$, but with about half of the computational time of $T=80$. $T=20$ gives relatively large deviation at scales smaller than $k=0.5\hMpc$. In Fig. \ref{fig:diffa0_fastpm}, we vary the initial redshift from 99, 19 and 9, respectively, while fixing $B=2$ and $T=40$. A larger initial redshift gives a more accurate IC from 2LPT, but results in a larger step size given a fixed number of time-steps. It turns out that $a_0=0.05$ gives better performance compared with the other two.

Based on the above tests, we choose our configuration parameters as $B=2$, $T=40$, and $a_0=0.05$. In fact, given a set of parameters, the performance would vary depending on redshifts. It is not guaranteed that the parameters we choose give better performance at some low redshifts than other choices. However, we do not strictly examine which set of parameters are the best, but try to find a reasonable one based on our goals.

\section{The effect from mass cut, abundance matching, and halo cleaning}\label{appendix:mc_am_hc}
\begin{figure}
    \centering
    \includegraphics[width=1.0\linewidth]{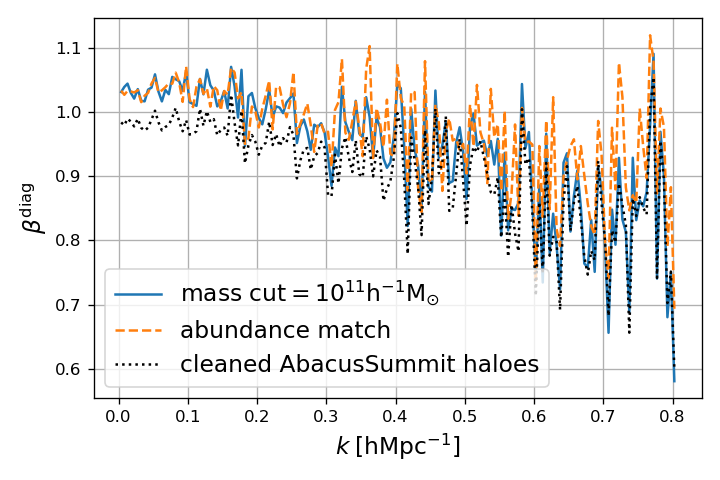}
    \caption{The comparison of $\beta^{\text{diag}}$ of the power spectrum monopoles calculated from the case of mass cut equal to $10^{11}\Msunh$ versus the case of abundance matching. For the abundance matching, we match the number of \textsc{FastPM} haloes close to that of \textsc{AbacusSummit} with mass cut. In addition, we overplot the result (black dotted line) from the case if we use the cleaned version of \textsc{AbacusSummit} halo catalogues.}
    \label{fig:p0_beta_mc_ab}
\end{figure}

Given a halo mass cut, e.g. $10^{11}\Msunh$, the halo number density of \textsc{FastPM} is about $36\%$ lower than that of \textsc{AbacusSummit} (uncleaned). To check whether the difference of number densities will reduce the CARPool performance or not, we select high-mass haloes from the paired \textsc{FastPM} catalogues to match with the number of haloes from \textsc{AbacusSummit} with mass cut. We compare the control variant $\beta^{\text{diag}}$ of the power spectrum monopoles from mass cut and abundance matching, and show the result in Fig. \ref{fig:p0_beta_mc_ab}. The improved performance of abundance matching (orange dashed line) is not very significant compared with that of mass cut (blue solid line). We have checked the conclusion is true for the quadrupole too. Furthermore, in the case of mass cut, we replace the uncleaned \textsc{AbacusSummit} halo catalogues by the cleaned ones and obtain the black dotted line. On large scales $k<0.2\Mpch$, $\beta^{\text{diag}}$ seems performing better than that from the uncleaned haloes as it is closer to 1.0. We expect that using the cleaned \textsc{AbacusSummit} haloes will benefit the CARPool performance on large scales.

\begin{figure}
    \centering
    \includegraphics[width=1.0\linewidth]{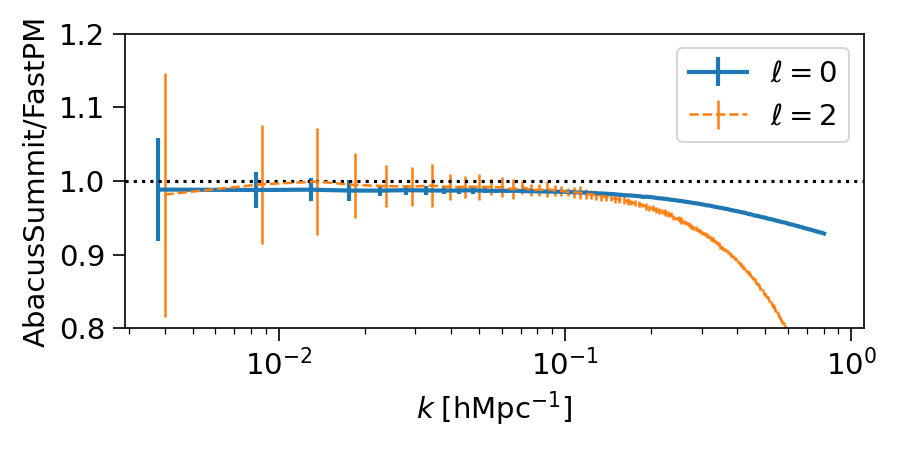}
    \caption{The ratio of the mean power spectrum monopoles ($\ell=0$) and quadrupoles ($\ell=2$) between the cleaned \textsc{AbacusSummit} and \textsc{FastPM} catalogues with mass cut $10^{11}\Msunh$. We slightly shift the $k$ coordinates of the quadrupole for clarity.}
    \label{fig:p02_cleaned_abacus}
\end{figure}

In Fig. \ref{fig:p02_cleaned_abacus}, we show the ratio of the power spectrum multipoles between the cleaned \textsc{AbacusSummit} and \textsc{FastPM} catalogues with mass cut $10^{11}\Msunh$. Compared with Fig. \ref{fig:pk_mcut1.e11}, the difference of halo biases at large scales between \textsc{AbacusSummit} and \textsc{FastPM} reduces significantly from $4.2\%$ to $1.2\%$ after the halo cleaning in \textsc{AbacusSummit}.

\section{CARPool performance of LRG-host haloes at redshift $0.8$}\label{appendix:z0.8}
\begin{figure}
\centering
    \includegraphics[width=1.0\linewidth]{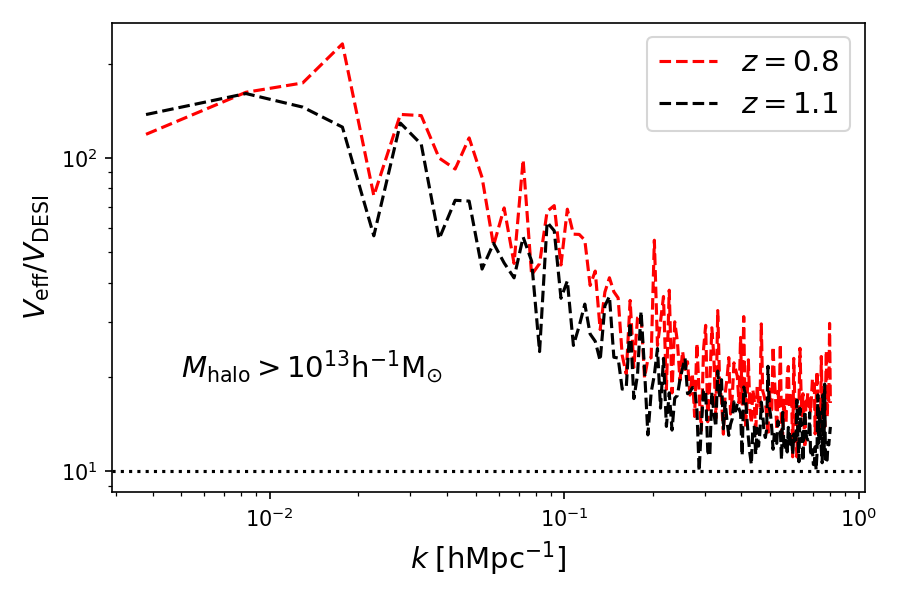}
    \caption{The comparison of the effective volumes obtained from CARPool for the halo power spectrum monopoles at redshifts $0.8$ and $1.1$. The haloes are massive than $10^{13}\Msunh$.}
    \label{fig:veff_p0_z0.8_1.1}
\end{figure}

For DESI LRGs, the number density peaks around redshift $0.6$--$0.8$ \citep{Zhou2021}. We simply check the performance of CARPool for the LRG-host haloes (with mass larger than $10^{13}\Msunh$) at $z=0.8$ compared with that from $z=1.1$. The number density of \textsc{AbacusSummit} haloes increases from $1.7\times 10^{-4}$ to $2.4\times 10^{-4}
\hMpccube$ from $z=1.1$ to $0.8$. We compare the effective volumes from CARPool at the two redshifts in Fig. \ref{fig:veff_p0_z0.8_1.1}. We show the results with the surrogate mean from the fixed-amplitude \textsc{FastPM} catalogues. Overall there is an average of $\sim 35\%$ increase on $V_{\text{eff}}$ at $z=0.8$, thanks to the increase of the halo number density. Here we use the surrogate mean from the fixed-amplitude \textsc{FastPM} catalogues. The increase is similar for the case using the surrogate mean from the non-fixed-amplitude catalogues.

\section{The Pearson correlation coefficients of the halo power spectrum multipoles between \textsc{AbacusSummit} and \textsc{FastPM}}\label{sec:cov_pell}
\begin{figure*}
    \centering
    \includegraphics[width=1.0\linewidth]{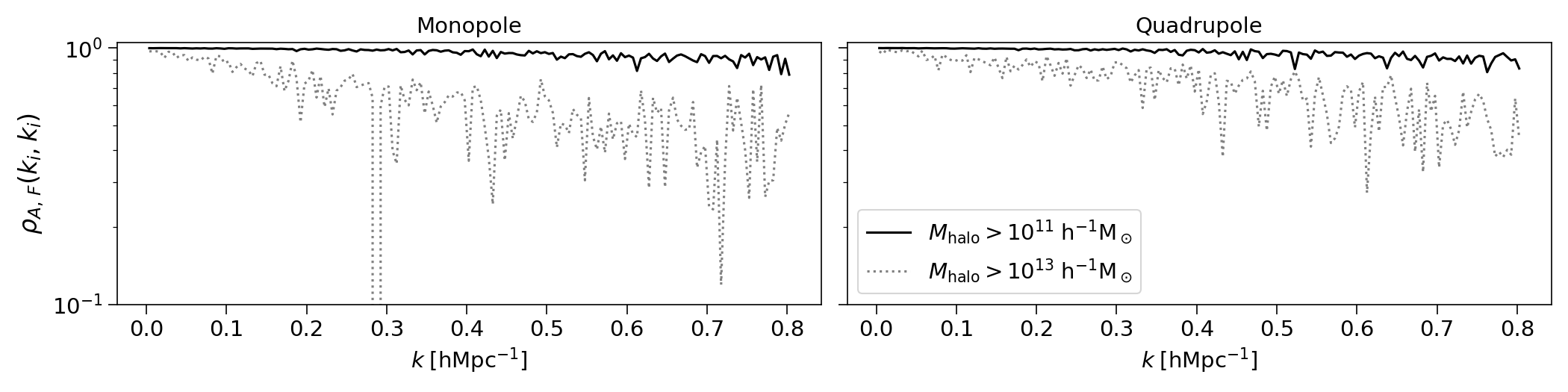}
    \caption{The diagonal terms of the Pearson correlation coefficients of the halo power spectrum multipoles between \textsc{AbacusSummit} and \textsc{FastPM}. The left-hand (right-hand) panel is for the monopole (quadrupole). The solid (dotted) lines are from the halo mass cut $10^{11}\Msunh$ ($10^{13}\Msunh$).}
    \label{fig:coeff_p0p2_ab_fp_1e11}
\end{figure*}

To compare the performance of CARPool for the halo power spectrum multipoles with different halo mass cuts ($10^{11}$ and $10^{13}\Msunh$) at $z=1.1$, we check the Pearson correlation coefficients between \textsc{AbacusSummit} and \textsc{FastPM}. For each mass cut, we calculate the power spectrum multipoles and the covariance matrices for \textsc{AbacusSummit} and \textsc{FastPM} with the matched \textsc{AbacusSummit} ICs.
We also calculate the cross-covariance matrix of the multipoles from the two simulations, and obtain the Pearson correlation coefficients via
\begin{align}
\rho_{\text{A},\, \text{F}}(k_i,\, k_j) = \frac{\text{Cov}_{\text{A},\, \text{F}}(k_i,\, k_j)}{\sqrt{\text{Cov}_{\text{A}}(k_i,\, k_i) \text{Cov}_{\text{F}}(k_j,\, k_j)}},
\end{align}
where the subscripts A and F denote \textsc{AbacusSummit} and \textsc{FastPM}, respectively.
In Fig. \ref{fig:coeff_p0p2_ab_fp_1e11}, we show the diagonal terms of $\rho_{ij}$ for the monopole (left-hand panel) and quadrupole (right-hand panel) with two mass cuts.
For both monopole and quadrupole, the Pearson coefficients are closer to $1.0$ from the halo mass cut $10^{11}\Msunh$, which indicates better performance from CARPool.

\section{Statistic noise of the correlation function quadrupole}\label{section:appendix_xi2}
We investigate the reason why there is about $2 \sim 3\sigma$ bias between the mean of the correlation function quadrupoles from the \textsc{AbacusSummit} halo catalogues and that from CARPool around $s=50\Mpch$ in Fig. \ref{fig:xix_mcut1e11}. 

\begin{figure}
    \centering
    \includegraphics[width=1.0\linewidth]{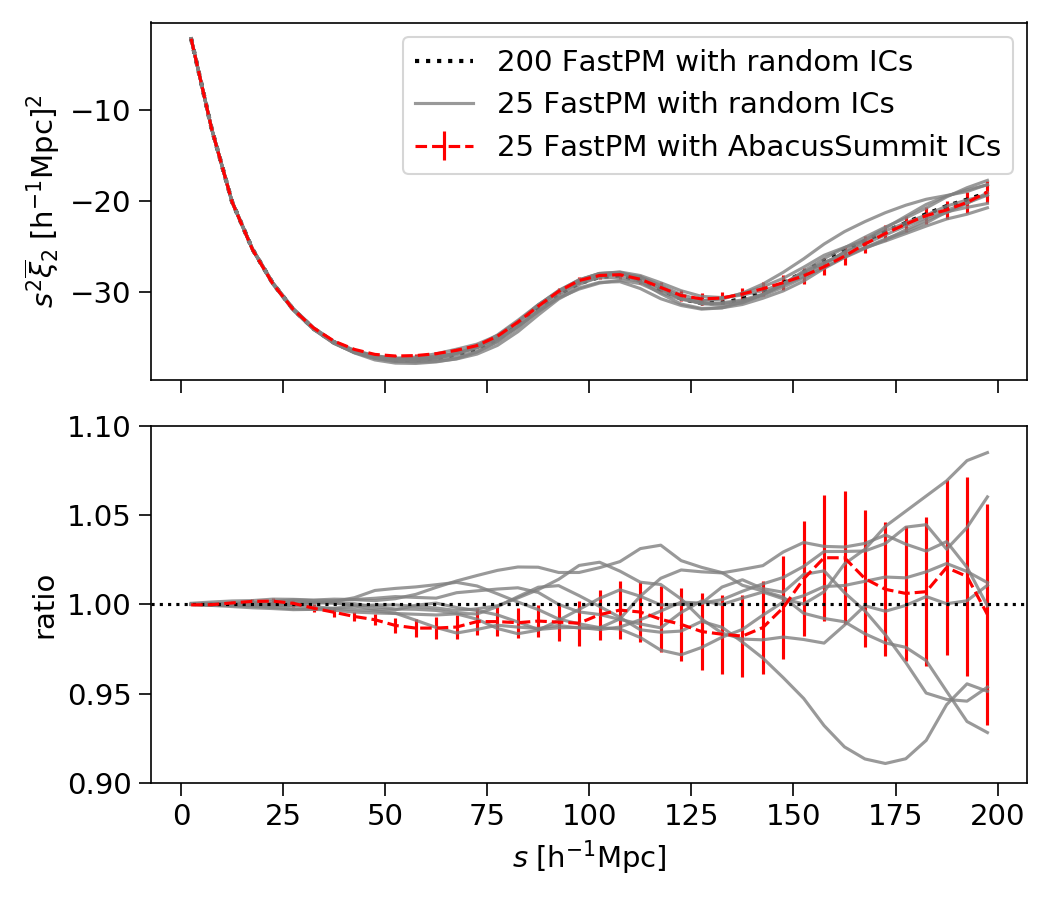}
    \caption{\textit{Upper panel}: we compare the correlation function quadrupoles from the \textsc{FastPM} halo catalogues with different cases. The dotted line shows the mean of the quadrupoles from 200 \textsc{FastPM} catalogues with random ICs. Each gray line represents the mean of every 25 out of 200 catalogues. The red dashed line is the mean of 25 \textsc{FastPM} catalogues with the \textsc{AbacusSummit} ICs. \textit{Lower panel}: we show the ratio of the result from each case by the mean of 200 \textsc{FastPM} catalogues. The red dashed line fluctuates around the horizontal dotted line and the fluctuation amplitude is similar to that of gray lines. At around $50$ $\Mpch$, there is about $2 \sim 3$ $\sigma$ deviation from 1.0, which is just due to the statistical noise.}
    \label{fig:xi2_fastpm_mcut1e11}
\end{figure}

We find that it is due to the statistical noise of the mean of the paired \textsc{FastPM} catalogues. In the upper panel of Fig. \ref{fig:xi2_fastpm_mcut1e11}, we show the mean of the quadrupoles from 200 \textsc{FastPM} halo catalogues with random ICs as the black dotted line. We divide these 200 catalogues into 8 groups, each of which has 25 catalogues. We calculate the mean for each group and plot it as a gray line. We also plot the mean of the paired \textsc{FastPM} catalogues with the \textsc{AbacusSummit} ICs as the red dashed line with error bars. In the lower panel, we divide all the results by the mean of 200 catalogues. We see that the fluctuation amplitude of the red dashed line is comparable with that of the gray lines. The deviation of the red dashed line from the black dotted line around $s=50\Mpch$ is about $2\sim 3\sigma$. Based on CARPool, we have the ratio of the mean between the \textsc{AbacusSummit} catalogues and that from CARPool as  
\begin{align}
    \frac{\overline{y}}{\overline{x}} = 1 + \beta \frac{\overline{c}-\hat{\mu}_c}{\overline{x}}, 
\end{align}
where $\overline{c}$ and $\hat{\mu}_c$ correspond to the red dashed line and black dotted line, respectively. Since we have checked that $\beta$ is close to 1.0 for $s>20\Mpch$, the statistical bias between $\overline{c}$ and $\hat{\mu}_c$ directly relates to the bias between the mean of \textsc{AbacusSummit} and CARPool.

\section{Cross-correlation between the clustering from different cosmologies}\label{section:appendix_cross_correlation}
\begin{figure*}
    \centering
    \includegraphics[width=1\linewidth]{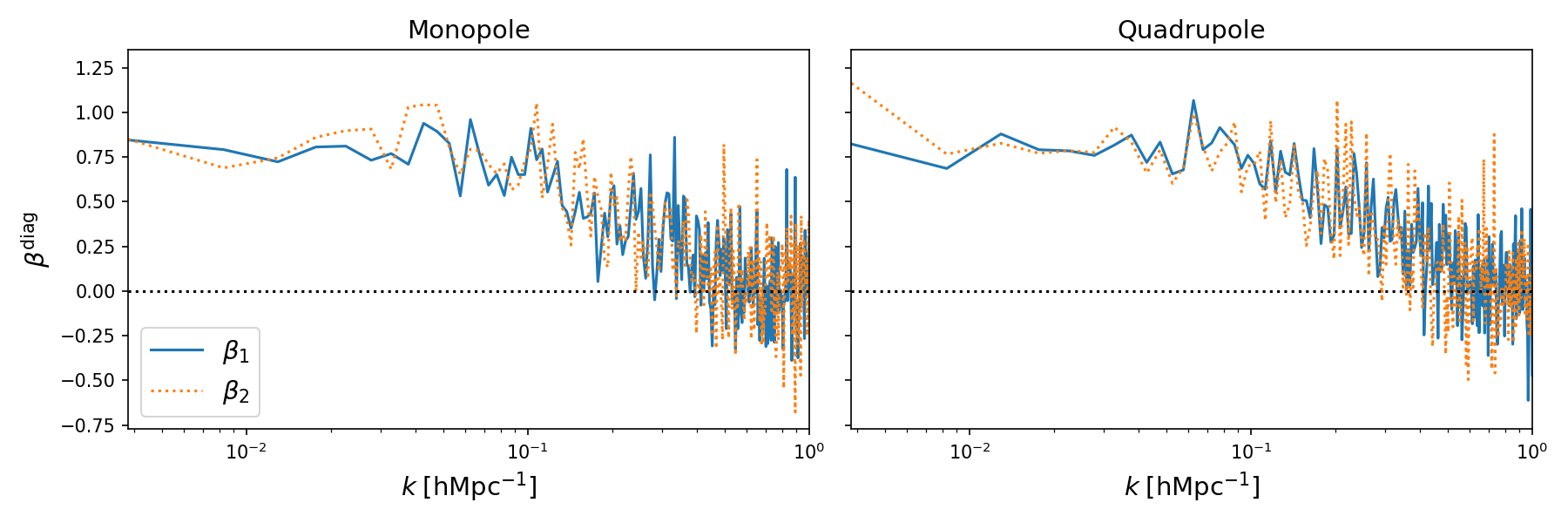}
    \caption{The shape of $\beta^{\text{diag}}$ which is calculated from the cross-correlation of the halo power spectrum multipoles between the cat2 and cat13 sharing the same ICs but with different cosmologies. $\beta_1$ is calculated from the cross-correlation between the cat2 with the cosmology c002 and the cat13 with the cosmology c000, while $\beta_2$ is calculated from the cat2 with c000 and the cat13 with c002. They agree well on the general shape for both monopole and quadrupole.}
    \label{fig:beta_cat2_cat13}
\end{figure*}

To test the validity of equation (\ref{eq:cov_c002_c000}), we use the halo catalogues of 25 paired \textsc{FastPM} simulations which are from the two cosmologies c000 and c002. Each paired simulation shares the same IC. We use the halo catalogues at $z=1.1$ with mass larger than $10^{11}\Msunh$. We divide haloes into three groups based on halo mass, i.e. $10^{11} < M_{\text{halo}} < 10^{12}\Msunh$, $10^{12} < M_{\text{halo}} < 10^{13}\Msunh$, and $M_{\text{halo}} > 10^{13}\Msunh$. We use cat1, cat2 and cat3 to represent them. As is known that the number of haloes decreases as the halo mass increases. We conduct subsampling for cat1 and cat2 with the percentage $2.85\%$ and $28\%$, respectively. The number of haloes after subsampling is about $7.2$ million for cat2. At the end, we reach two goals: one is that after subsampling, the number of the haloes from the combined cat1 and cat3, denoted as cat13, is close to that of cat2; the other is that the halo clustering statistics is similar between cat2 and cat13. Doing such process, we can mimic cat2 as a catalogue from an $N$-body simulation and cat13 as a catalogue from a paired surrogate, since cat13 is constructed to mimic cat2 in terms of the shot noise and two-point clustering signal and it shares the same IC with cat2. We obtain cat2 and cat13 from 25 halo catalogues in each cosmology. We calculate the halo power spectrum multipoles from cat2 and cat13, and compare $\beta_1$ and $\beta_2$ calculated from the cross-correlation and variance of the multipoles over 25 realizations, i.e.
\begin{align}
    \beta_1 = \frac{\text{Diag}[\text{Cov}(\text{cat2}_{\text{c}002}, \text{cat13}_{\text{c}000})]}{\text{Var}(\text{cat13}_{\text{c}000})},\\
    \beta_2 = \frac{\text{Diag}[\text{Cov}(\text{cat2}_{\text{c}000}, \text{cat13}_{\text{c}002})]}{\text{Var}(\text{cat13}_{\text{c}000})}.
\end{align}
We show the results in Fig. \ref{fig:beta_cat2_cat13}, in which the blue lines denote $\beta_1$ and the orange dotted lines denote $\beta_2$. They have a similar shape for both monopole and quadrupole. 

\section{CARPool results of the secondary cosmology c004}
\begin{figure}
    \centering
    \includegraphics[width=1.0\linewidth]{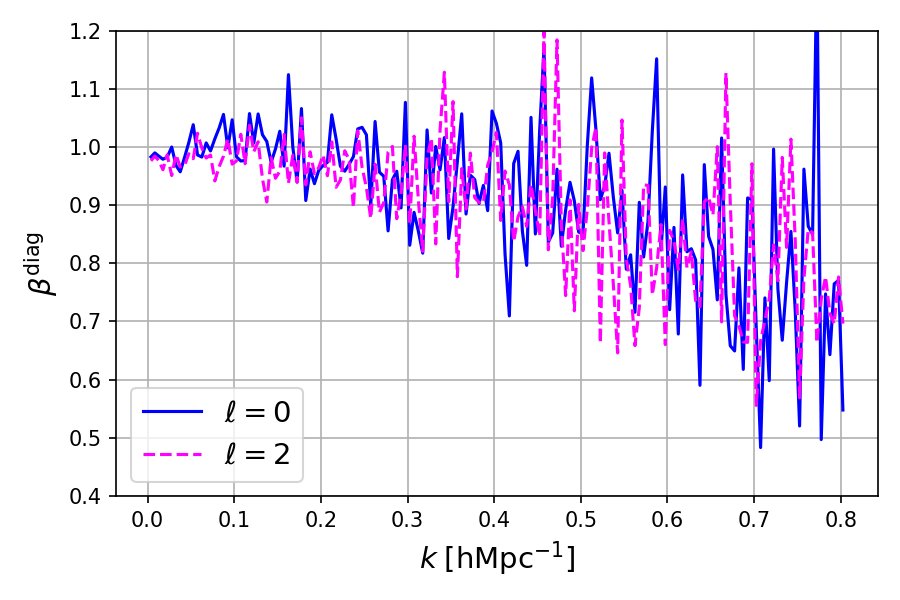}
    \caption{$\beta^{\text{diag}}$ is calculated from the cross-correlation between the halo power spectrum multipoles from the paired \textsc{AbacusSummit} in c000 and the \textsc{FastPM} in c004. $\ell=0\, (2)$ is for the monopole (quadrupole).}\label{fig:beta_c000_c004}
\end{figure}

\begin{figure*}
    \centering
    \includegraphics[width=1.\linewidth]{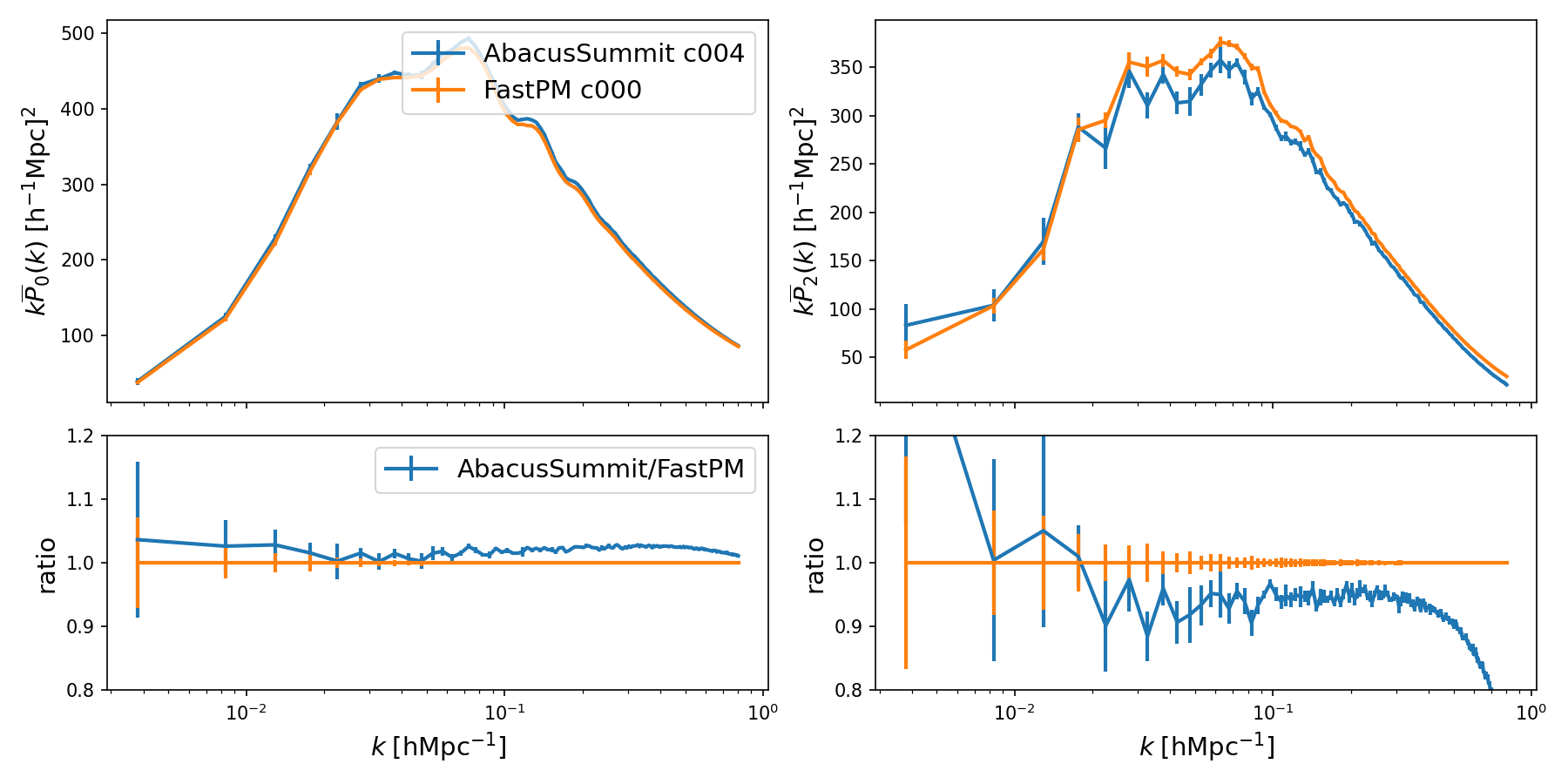}
    \caption{Same as Fig. \ref{fig:pk_c002} but for the secondary cosmology c004.}
    \label{fig:pk_c004}
\end{figure*}

\begin{figure*}
    \centering
    \includegraphics[width=1.0\linewidth]{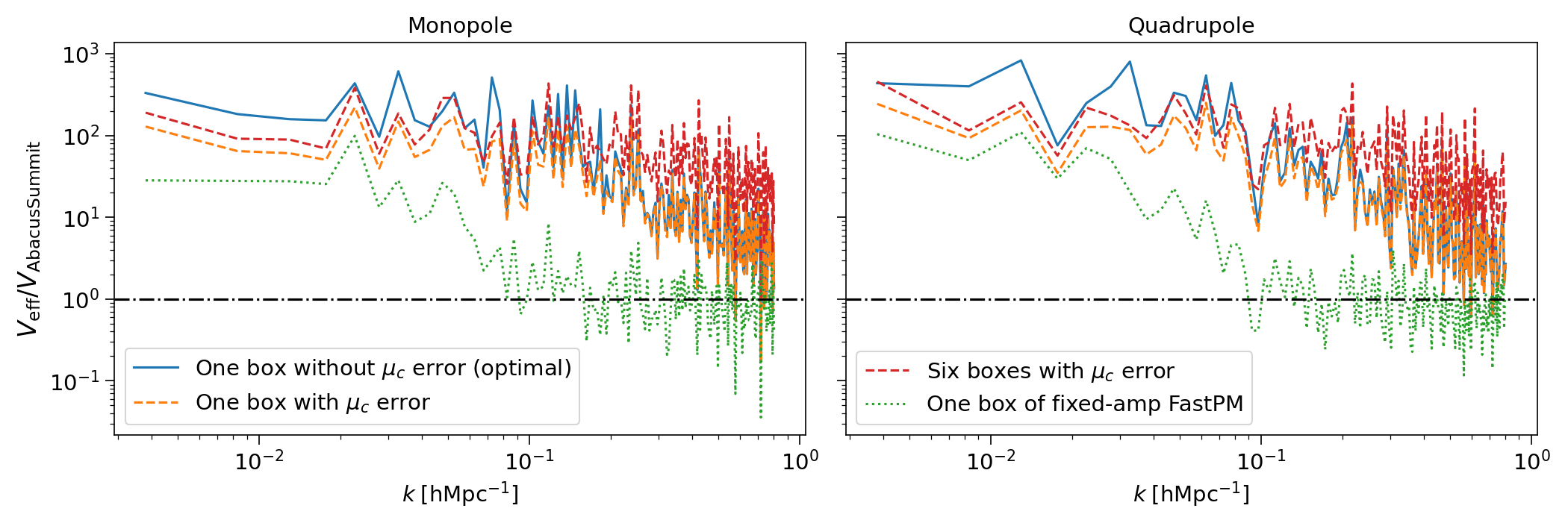}
    \caption{Same as Fig. \ref{fig:veff_c002} but for the secondary cosmology c004.}
    \label{fig:veff_c004}
\end{figure*}

Similar to Section \ref{sec:diff_cosmo}, we apply CARPool on the halo catalogues with the secondary cosmology c004. We compare the halo power spectrum multipoles from c004 and c000 in Fig. \ref{fig:pk_c004}. Compared with Fig. \ref{fig:pk_c002}, we can see that the difference of the power spectrum monopoles between c004 and c000 is smaller than that between c002 and c000, which is expected as c004 is different from c000 only on $\sigma_8$. We show the $\beta$ in Fig. \ref{fig:beta_c000_c004}, and the effective volume increased from CARPool in Fig. \ref{fig:veff_c004}.

\label{appendix:carpool_c004}

\bsp	
\label{lastpage}
\end{document}